\documentclass[oldversion]{aa}  
\usepackage{graphicx}
\usepackage{natbib}

\bibpunct{(}{)}{;}{a}{}{,}

\def\lsim{\mathrel{\hbox{\rlap{\lower.55ex \hbox {$\sim$}}\kern-.0em
\raise.4ex \hbox{$<$}}}} 
\def\gsim{\mathrel{\hbox{\rlap{\lower.55ex \hbox {$\sim$}}\kern-.0em
\raise.4ex \hbox{$>$}}}} 
\def\lya{Ly$\alpha$}

\def\ion#1#2{#1$\;${\small\rm\@Roman{#2}}\relax}
\def\la{$\lambda$}
\def\kms{km s$^{-1}$}
\def\1star{$^{\star}$}
\def\2star{$^{\star\star}$}
\def\3star{$^{\star\star\star}$}
\def\4star{$^{\star\star\star\star}$}
\def\a{$^{\rm a}$}
\def\b{$^{\rm b}$}

\def\abs{$_{\rm abs}$}
\def\grb{GRB\,060418}
\newcommand{\gpm}[3]{$#1^{+#2}_{-#3}$}
\newcommand{\swift}{{\it Swift}}

\begin{document}

\title{Rapid-Response Mode VLT/UVES spectroscopy of \grb\thanks{Based
    on observations collected at the European Southern Observatory,
    Chile; proposal no. 77.D-0661.}}

\subtitle{Conclusive evidence for UV pumping from the time evolution
  of Fe~II and Ni~II excited- and metastable-level populations}

\author{P.~M. Vreeswijk\inst{1,2}
  \and
  C. Ledoux\inst{1}
  \and
  A. Smette\inst{1}
  \and
  S.~L. Ellison\inst{3}
  \and
  A.~O. Jaunsen\inst{4}
  \and
  M.~I. Andersen\inst{5}
  \and
  A.~S. Fruchter\inst{6} 
  \and
  J.~P.~U. Fynbo\inst{7}
  \and
  J. Hjorth\inst{7}
  \and
  A. Kaufer\inst{1}
  \and
  P. M{\o}ller\inst{8}
  \and
  P. Petitjean\inst{9,10}
  \and
  S. Savaglio\inst{11}
  \and
  R.~A.~M.~J. Wijers\inst{12}
}  


\institute{European Southern Observatory, Alonso de C\'ordova 3107,
  Casilla 19001, Santiago 19, Chile
  \and
  Departamento de Astronom\'ia, Universidad de Chile, Casilla 36-D,
  Santiago, Chile
  \and
  Department of Physics and Astronomy, University of Victoria,
  Victoria, BC, Canada
  \and
  Institute of Theoretical Astrophysics, University of Oslo, PO Box
  1029 Blindern, 0315 Oslo, Norway
  \and
  Astrophysikalisches Institut Potsdam, An der Sternwarte 16, D-14482
  Potsdam, Germany
  \and
  Space Telescope Science Institute, 3700 San Martin Drive, Baltimore,
  MD 21218, USA
  \and
  Dark Cosmology Centre, Niels Bohr Institute, University of
  Copenhagen, DK-2100 Copenhagen, Denmark
  \and
  European Southern Observatory, Karl-Schwarzschild-Str. 2,
  D-85748, Garching bei M\"unchen, Germany
  \and
  Institut d'Astrophysique de Paris - UMR 7095 CNRS \& Universit\'e
  Pierre et Marie Curie, 98bis Boulevard Arago, 75014 Paris, France
  \and
  LERMA, Observatoire de Paris, 61 Avenue de l'Observatoire, 75014
  Paris, France
  \and
  Max-Planck-Institut f\"ur Extraterrestrische Physik,
  Giessenbachstrasse, D-85748 Garching bei M\"unchen, Germany
  \and
  Astronomical Institute `Anton Pannekoek', University of Amsterdam \&
  Center for High Energy Astrophysics, Kruislaan 403, 1098 SJ
  Amsterdam, The Netherlands 
}

\date{\today}

\authorrunning{Vreeswijk, Ledoux, Smette et al.}
\titlerunning{Rapid-Response Mode VLT/UVES spectroscopy of \grb}


\abstract{We present high-resolution spectroscopic observations of
  \grb, obtained with VLT/UVES. These observations were triggered
  using the VLT Rapid-Response Mode (RRM), which allows for automated
  observations of transient phenomena, without any human
  intervention. This resulted in the first UVES exposure of \grb\ to
  be started only 10 minutes after the initial \swift\ satellite
  trigger. A sequence of spectra covering 330-670~nm were acquired at
  11, 16, 25, 41 and 71 minutes (mid-exposure) after the trigger, with
  a resolving power of 7~\kms, and a signal-to-noise ratio of 10-15.
  This time-series clearly shows evidence for time variability of
  allowed transitions involving Fe~II fine-structure levels
  ($^6$D$_{7/2}$, $^6$D$_{5/2}$, $^6$D$_{3/2}$, and $^6$D$_{1/2}$),
  and metastable levels of both Fe~II ($^4$F$_{9/2}$ and
  $^4$D$_{7/2}$) and Ni~II ($^4$F$_{9/2}$), at the host-galaxy
  redshift $z=1.490$. This is the first report of absorption lines
  arising from metastable levels of Fe~II and Ni~II along any GRB
  sightline. We model the observed evolution of the level populations
  with three different excitation mechanisms: collisions, excitation
  by infra-red photons, and fluorescence following excitation by
  ultraviolet photons. Our data allow us to reject the collisional and
  IR excitation scenarios with high confidence. The UV pumping model,
  in which the GRB afterglow UV photons excite a cloud of atoms with a
  column density $N$, distance $d$, and Doppler broadening parameter
  $b$, provides an excellent fit, with best-fit values:
  log~$N$(Fe~II)=\gpm{14.75}{0.06}{0.04},
  log~$N$(Ni~II)=$13.84\pm0.02$, $d=1.7\pm0.2$~kpc (but see Appendix
  A), and $b=25\pm3$~\kms. The success of our UV pumping modeling
  implies that no significant amount of Fe~II or Ni~II is present at
  distances smaller than $\sim$1.7~kpc (but see erratum in Appendix
  A), most likely because it is ionized by the GRB X-ray/UV
  flash. Because neutral hydrogen is more easily ionized than Fe~II
  and Ni~II, this minimum distance also applies to any H~I present.
  Therefore the majority of very large H~I column densities typically
  observed along GRB sightlines may not be located in the immediate
  environment of the GRB.  The UV pumping fit also constrains two GRB
  afterglow parameters: the spectral slope,
  $\beta=$\gpm{-0.5}{0.8}{1.0}, and the total rest-frame UV flux that
  irradiated the cloud since the GRB trigger, constraining the
  magnitude of a possible UV flash.}


\maketitle
%

\begin{table*}[t]
  \centering
  \caption[]{Log of UVES observations}\label{tab:log}
  \null\vspace{-1.0cm}
  $$
  \begin{array}{cccrclrcccr}
    \hline
    \hline
    \noalign{\smallskip}
    \rm UT \, start &
    \rm epoch &
    \rm Fig.~\ref{fig:profiles}~line~colour&
    \rm \Delta T^{a} &
    \lambda_{\rm central} &
    \rm coverage &
    \rm exptime &
    \rm seeing^{b} &
    \rm FWHM &
    \rm S/N &
    \rm OT\,mag^{c} \\
    \rm (2006 \, April \, 18) &
    &
    &
    \rm (min) &
    \rm (nm) &
    \rm (nm) &
    \rm (min) &
    \rm (\arcsec) &
    ($\kms$) &
    \rm peak &
    \\
    \hline
$3:16:07$ & 1 & \rm black   & 11.47 &  390 &  328$-$452              & 3  & 1.5  & 6.9 & 6 &\rm R~(6min)=14.0\\
$3:16:02$ & 1 &             & 11.38 &  564 &  462$-$560 ; 568$-$665  & 3  & 1.1  & 7.2 & 14&\rm V~(6.5min)=15.0\\
$3:20:17$ & 2 & \rm red     & 16.61 &  390 &  328$-$452              & 5  & 1.5  & 6.9 & 6 &\rm z~(16.3min)=14.4 \\
$3:20:12$ & 2 &             & 16.52 &  564 &  462$-$560 ; 568$-$665  & 5  & 1.0  & 7.2 & 15&     \\
$3:26:27$ & 3 & \rm blue    & 25.13 &  390 &  328$-$452              & 10 & 1.5  & 6.9 & 7 &     \\
$3:26:17$ & 3 &             & 24.96 &  564 &  462$-$560 ; 568$-$665  & 10 & 1.0  & 7.2 & 16&     \\
$3:38:01$ & 4 & \rm green   & 41.20 &  390 &  328$-$452              & 20 & 1.4  & 6.9 & 7 &     \\
$3:37:56$ & 4 &             & 41.12 &  564 &  462$-$560 ; 568$-$665  & 20 & 0.9  & 7.2 & 17&     \\
$3:59:30$ & 5 & \rm magenta & 70.99 &  390 &  328$-$452              & 40 & 1.4  & 6.9 & 7 &\rm I~(69min)=16.5\\
$3:59:24$ & 5 &             & 70.88 &  564 &  462$-$560 ; 568$-$665  & 40 & 0.9  & 7.2 & 17&\rm z~(78min)=16.2\\
$4:41:51$ & 6 &             & 128.09 & 437 &  376$-$498              & 80 & 1.0  & 6.1 & 9 &\rm V~(100min)=18.8\\
$4:41:46$ & 6 &             & 128.00 & 860 &  670$-$852 ; 866$-$1043 & 80 & 0.8  & 6.5 & 21&\rm I~(135min)=17.4\\
    \hline
  \end{array}
  $$ \flushleft 
  $^{\rm a}$ Time of flux-weighted mid-exposure since GRB trigger,
  assuming the light curve decay index $\alpha=-1.1$.\\
  $^{\rm b}$ The seeing has been estimated from the 2-D spectra.\\
  $^{\rm c}$ Approximate magnitude of the optical transient around the
  time of our spectra, in filters V \citep{2006GCN..4978....1S}, R
  \citep{Melandri06_GCN4968}, I \citep{Cobb06_GCN4972} or z
  \citep{Nysewander06_GCN4971}.\\
\end{table*}

\section{Introduction}
\label{sec:introduction}

The influence of a $\gamma$-ray burst (GRB) explosion on its
environment has been predicted to manifest itself in various
ways. Strong observational evidence
\citep{1998Natur.395..670G,stanek,hjorth030329} indicates that at
least some GRB progenitors are massive stars
\citep{1993ApJ...405..273W,1999ApJ...524..262M}, and therefore the
explosion is likely to take place in a star-forming region. As the GRB
radiation ionizes the atoms in the environment, the neutral hydrogen
and metal column densities in the vicinity of the explosion are
expected to evolve with time
\citep{1998ApJ...501..467P,2001ApJ...546..672V,2002ApJ...578..818M}.
Ultra-violet (UV) photons will not only photo-dissociate and
photo-ionize any nearby molecular hydrogen, but also quickly excite
H$_2$ at larger distances to its vibrationally excited metastable
levels, which can be observed in absorption
\citep{2002ApJ...569..780D}. Finally, dust grains can be destroyed up
to tens of parsecs away
\citep{2000ApJ...537..796W,2001ApJ...563..597F,2002ApJ...569..780D,2002ApJ...580..261P,2003ApJ...585..775P}. Detection
of these time-dependent processes, with timescales ranging from
seconds to days in the observer's frame, would not only provide direct
information on the physical conditions of the interstellar medium
(ISM) surrounding the GRB, but would also constrain the properties of
the emitted GRB flux before it is attenuated by foreground absorbers
in the host galaxy and in intervening gas clouds. In the X-ray,
evidence has been found for a time-variable H~I column density
\citep{2005A&A...442L..21S,2006astro.ph.11305C}, presumably due to the
ionization of the nearby neutral gas. In the optical, none of these
processes have been observed until recently, when
\citet{2006astro.ph..6462D} reported a $\sim$3$\sigma$ variability
detection of Fe~II $^6$D$_{7/2}$ \la 2396\footnote{Lines arising from
  fine-structure levels are sometimes indicated with stars, e.g.
  Fe~II\1star\ for Fe~II $^6$D$_{7/2}$, Fe~II\2star\ for Fe~II
  $^6$D$_{5/2}$, etc.; in this paper we will instead list the
  transition lower energy level term and J value in order to indicate
  all levels that we will discuss in a consistent manner.}, observed
at two epochs roughly 16 hours apart. Such observations are
technically very challenging because high-resolution spectroscopy
combined with the rapidly decaying afterglow flux requires immediate
follow-up with 8-10~m class telescopes.

The \swift\ satellite, launched in November 2004, has permitted a
revolution in rapid spectroscopic follow-up observations, providing
accurate (5\arcsec) positions for the majority of GRBs within a few
minutes of the GRB trigger. Numerous robotic imaging telescopes react
impressively fast (within 10~sec) to \swift\ triggers.  As for
spectroscopic observations, a number of target-of-opportunity programs
at most major observational facilities are regularly yielding
follow-up observations of the GRB afterglow at typically an hour after
the \swift\ alert.  However, most of these programs require
significant human coordination between the science team and telescope
personnel/observers.  At the European Southern Observatory's (ESO)
Very Large Telescope (VLT; consisting of four unit telescopes of 8.2~m
each), a Rapid Response Mode (RRM) has been commissioned to provide
prompt follow-up of transient phenomena, such as GRBs.  The design of
this system\footnote{see http://www.eso.org/observing/p2pp/rrm.html}
allows for completely automatic VLT observations without any human
intervention except for the placement of the spectrograph entrance
slit on the GRB afterglow. The typical time delay, which is mainly
caused by the telescope preset and object acquisition, is 5-10
minutes, depending on the GRB location on the sky with respect to the
telescope pointing position prior to the GRB alert. The data presented
in this paper are the result of the first automatically-triggered RRM
activation.

This paper is organized as follows: the UVES observations and data
reduction are described in Sect.~\ref{sec:observations}, followed by
Sect.~\ref{sec:hostabsorbers}, in which we discuss general properties
of the absorption systems at the host-galaxy redshift from the
detection of resonance lines. In Sect.~\ref{sec:variability}, we focus
on the detection of variability of transitions originating from
fine-structure levels of Fe~II, and metastable levels of Fe~II and
Ni~II. The time evolution of the level population of these excited
levels is modeled in Sect.~\ref{sec:modeling}. The results and their
implications are discussed in Sect.~\ref{sec:discussion}, and we
conclude in Sect.~\ref{sec:conclusions}.

\section{UVES observations and data reduction}
\label{sec:observations}

On April 18 2006 at 3:06:08 UT the \swift\ Burst Alert Telescope (BAT)
triggered a $\gamma$-ray burst alert \citep{2006GCN..4966....1F},
providing a 3\arcmin\ error circle localization. Observations with the
\swift\ X-Ray Telescope (XRT) resulted in a 5\arcsec\ position about
one minute later \citep{2006GCN..4973....1F}, which triggered our
desktop computer to activate a VLT-RRM request for observations with
the Ultra-violet and Visual Echelle Spectrograph (UVES). This was
received by the VLT's unit telescope Kueyen at Cerro Paranal at
3:08:12 UT.  The on-going service mode exposure was ended immediately,
and the telescope was pointed to the XRT location, all
automatically. Several minutes later, the night astronomers Stefano
Bagnulo and Stan Stefl identified the GRB afterglow, aligned the UVES
slit on top of it, and started the requested observations at 3:16 UT
(i.e. 10 minutes after the \swift\ $\gamma$-ray detection). This
represents the fastest spectral follow-up of any GRB by an optical
facility \citep[until the RRM VLT/UVES observations of GRB~060607,
  also triggered by our team, which were started at a mere 7.5 minutes
  after the GRB;][]{Ledoux06_GCN5237}. A series of exposures with
increasing integration times (3, 5, 10, 20, and 40 minutes,
respectively) was performed with a slit width of 1\arcsec, yielding
spectra covering the 330--670 nm wavelength range at a resolving power
of $R = \Delta \lambda / \lambda$ $\sim$ 43,000, corresponding to
7~\kms\ full width at half maximum. These observations were followed
by a 80-minutes exposure in a different instrument configuration, but
with the same slit width, extending the wavelength coverage to the red
up to 950~nm.  The data were reduced with a custom version of the UVES
pipeline \citep{ballester}, flux-calibrated using the standard
response
curves\footnote{http://www.eso.org/observing/dfo/quality/UVES/qc/\\
  response.html} and converted to a heliocentric vacuum wavelength
scale. The log of the observations is shown in Table~\ref{tab:log}.

\begin{table*}[htb]
  \centering 
  \caption {Ionic column densities and abundance ratios in the
    combined spectrum (see Fig.\ref{fig:sumprofiles} for the
    corresponding profile fits).
    \label{tab:groundlevels}}
  \null\vspace{-1.0cm}
  $$
  \begin{array}{llrrrr}
    \hline
    \hline
    \rm Ion & 
    \rm Lines\,\,used &
    \multicolumn{4}{c}{\log~N\pm\sigma _{\log~N}$\a$}\\
    \hline
    \rm component                     && 1            & 2            & 3            & \rm total \\
    z$\abs$                           && 1.49001(1)   & 1.48999(7)   & 1.49018(3)   & \\
    b~($\kms$)                        && 23.6\pm0.4   & 3.0\pm0.4    & 6.5\pm0.1    & \\
    \hline
    \rm C~I   & 1656                  &  12.52\pm0.18 & 12.29\pm0.14 & 12.19\pm0.17 & 12.83\pm0.11 \\
    \rm Mg~I  & 2026                  &  <12.07       & 13.01\pm0.02 & 13.60\pm0.01$\b$ & 13.70\pm0.02$\b$ \\
    \rm Si~II & 1808                  &  15.74\pm0.02 & 14.85\pm0.17 & 15.34\pm0.05 & 15.92\pm0.03 \\
    \rm Cr~II & 2056, 2062, 2066      &  13.51\pm0.01 & 12.46\pm0.06 & 12.88\pm0.03 & 13.63\pm0.02 \\
    \rm Mn~II & 2576, 2594, 2606      &  12.97\pm0.01 & 11.82\pm0.05 & 12.59\pm0.02 & 13.14\pm0.01 \\
    \rm Fe~II & 1611                  &  14.88\pm0.10 & 14.03\pm0.31 & 14.49\pm0.11 & 15.07\pm0.08 \\
    \rm Ni~II & 1709, 1741, 1751      &  13.77\pm0.03 & 12.82\pm0.13 & 13.03\pm0.08 & 13.88\pm0.03 \\
    \rm Zn~II & 2026, 2062            &  12.80\pm0.02 & 11.98\pm0.05 & 12.69\pm0.01 & 13.09\pm0.01 \\
    \hline
    \multicolumn{6}{l}{\rm Abundance\,\,ratio}\\
    \rm [Si/Fe]                      &&   0.79\pm0.10 &  0.75\pm0.35 &  0.78\pm0.12 &  0.78\pm0.09 \\
    \rm [Cr/Fe]                      &&   0.45\pm0.10 &  0.25\pm0.32 &  0.21\pm0.11 &  0.38\pm0.08 \\
    \rm [Mn/Fe]                      &&   0.06\pm0.10 & -0.24\pm0.31 &  0.07\pm0.11 &  0.04\pm0.08 \\
    \rm [Ni/Fe]                      &&   0.14\pm0.10 &  0.04\pm0.34 & -0.21\pm0.14 &  0.06\pm0.09 \\
    \rm [Zn/Fe]                      &&   0.76\pm0.10 &  0.79\pm0.31 &  1.04\pm0.11 &  0.86\pm0.08 \\
    \hline
  \end{array}
  $$
  \flushleft
  \a\ The errors listed are the formal errors provided by FITLYMAN;
  although an error of 0.02 on an individual component is likely to be
  an underestimate, we consider the error on the total column density
  to be realistic.\\
  \b\ The third component of Mg~I is likely to be blended; see the 
  discussion in the text.\\
\end{table*}

\begin{figure*}
  \centering
  \includegraphics[width=9.0cm,clip=true]{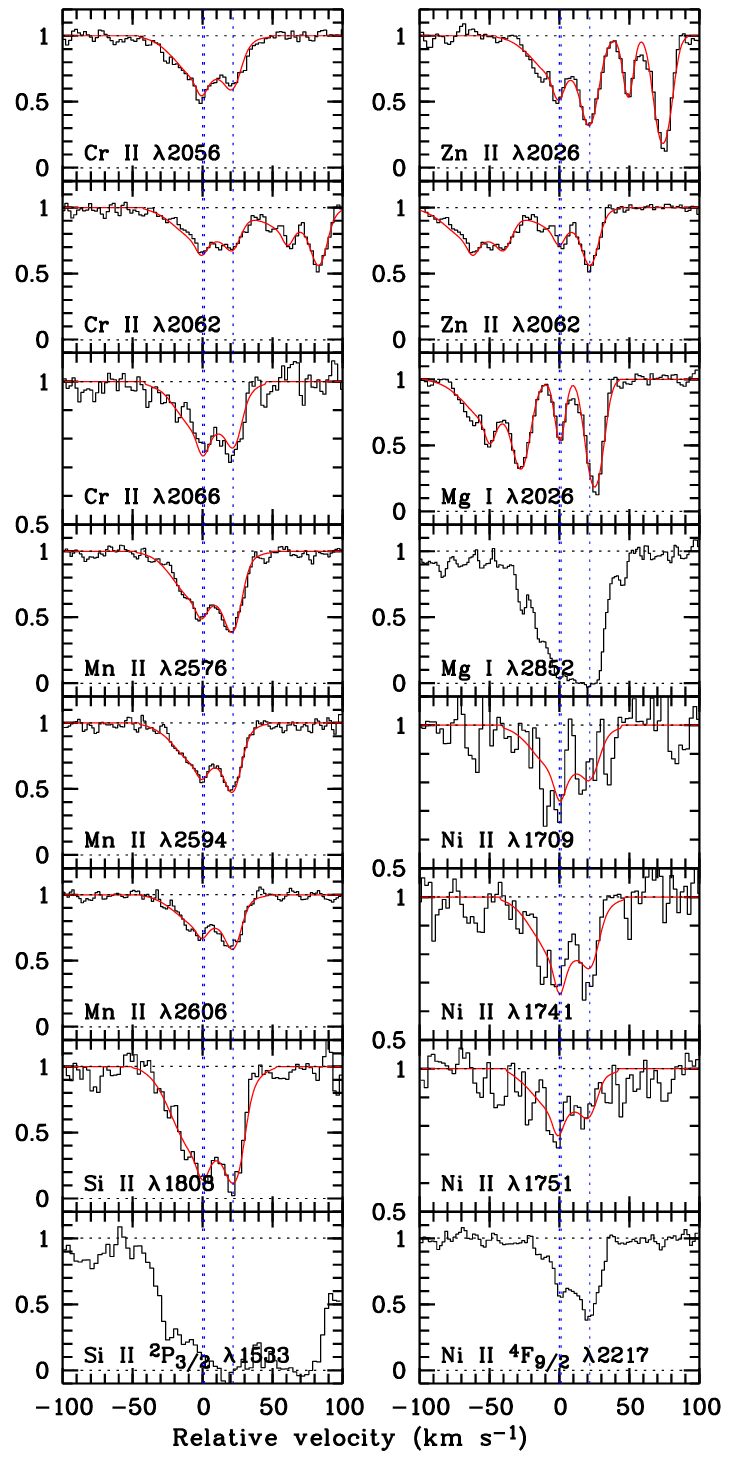}
  \includegraphics[width=9.0cm,clip=true]{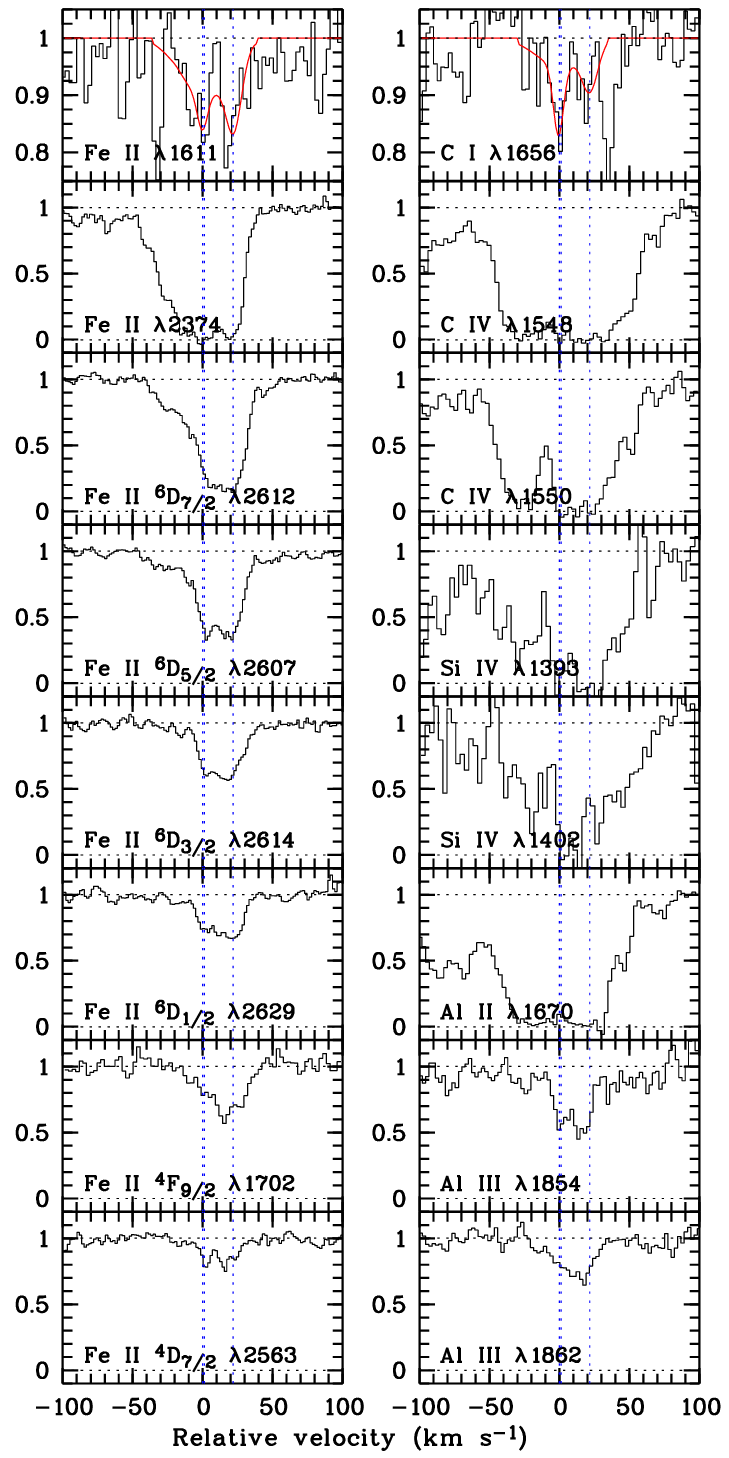}
  \caption{Absorption-line profiles for a variety of transitions
    detected at the \grb\ redshift. To all low-ionization species
    containing at least one non-saturated transition, we have
    performed simultaneous Voigt-profiles fits using a three-component
    model; the resulting fits are shown by the solid (red) line (see
    Table~\ref{tab:groundlevels} for the fit results). The relative
    velocity of the different components are indicated by the (blue)
    vertical dotted lines; note that two components have very similar
    redshifts. It is clear that this 3-component fit does not describe
    some high S/N lines, such as Zn~II \la 2026, very well; we will
    come back to this in the discussion. We note, however, that the
    total column density would hardly change if additional components
    would be introduced. For comparison purposes, we also show
    transitions that we have not fit: (saturated) higher-ionization
    lines of C~IV and Si~IV, lines from Fe~II fine-structure levels,
    and transitions originating from metastable levels of both Fe~II
    and Ni~II. \label{fig:sumprofiles}}
\end{figure*}

\section{Ground-state absorption lines from the host galaxy of \grb}
\label{sec:hostabsorbers}

The spectra reveal four strong absorption-line systems at redshifts $z
= 0.603, 0.656, 1.107$, and 1.490. In what follows, we focus on the
highest-redshift absorption-line system at $z_{\rm abs}=1.490$, which
corresponds to the redshift of the GRB as shown below; the intervening
systems are discussed in a separate paper \citep{sara060418}.

At the redshift of the GRB host galaxy, we detect a large number of
metal-absorption lines, arising from transitions involving the ground
state of various ions (see below), fine-structure levels of Si~II and
Fe~II, and metastable levels of Fe~II and Ni~II. We will discuss the
excited-level lines in more detail in Sect.~\ref{sec:variability}; in
this section we focus on the resonance lines, i.e. lines corresponding
to an allowed transition from the ion ground state to a higher excited
level.  The ions from which resonance lines are detected are C~I,
C~IV, Cr~II, Mn~II, Si~II, Si~IV, Zn~II, Mg~I, Mg~II, Ni~II, Fe~II,
Al~II, Al~III, and Ca~II.  C~II \la 1334 is at the very blue edge of
our spectrum, where the noise is dominating the signal. For Fe~I, we
determine an upper limit (5$\sigma$) on its column density of
log~N(Fe~I)$<$11.48, using Fe~I \la 2484. A selection of lines is shown
in Fig.~\ref{fig:sumprofiles}. Because the resonance lines are not
found to vary in time, we have combined all spectra (see
Table~\ref{tab:log}) to achieve the highest signal-to-noise ratio
(S/N) possible. The combined spectrum has S/N=16 at
\la=4000~\AA\ (\la$_{\rm rest}$=1606~\AA) and S/N=36 at
\la=6500~\AA\ (\la$_{\rm rest}$=2610~\AA).  The redshift corresponding
to the zero velocity has been adopted to be $z=1.49000$. At this
redshift, the \lya\ line is located at 3027~\AA, just outside the UVES
spectral range.

We now wish to highlight a number of observations that can be made
from Fig.~\ref{fig:sumprofiles}. The vast majority of the column
density of the low- and high-ionization species (as well as
fine-structure and metastable species) is located within a narrow
range of velocity (with a spread of 50-100~\kms), and seems to be
contained within two or three main components. This small range in
velocity for the low-ionization species is also seen in GRB~051111
\citep{2006astro.ph..1057P,2006astro.ph.11092P}. Highly saturated
lines such as those from C~IV, Si~IV, Mg~II, and Al~II show components
to the blue up to $-200$~\kms, but these harbour only a small fraction
of the total column density. The line profile of the high-ionization
C~IV lines follows the profile of the low-ionization lines very well,
even though the comparison is made difficult by the fact that the C~IV
lines are much stronger. This similarity is uncommon in QSO-DLAs
\citep{2000ApJ...545..591W}. The main clump of the low-ionization line
profiles, though kinematically simple, is a complex mix of broad and
narrow components. Thanks to the high signal-to-noise ratio, a weak
narrow component is clearly observed in the profiles of Cr~II and
Mn~II at $-20$~\kms.

For a quantitative analysis, we have simultaneously fit Voigt-profiles
(using the FITLYMAN context within MIDAS) to all resonance lines with
at least one non-saturated transition. The atomic data required by the
profile fits (vacuum wavelengths, oscillator strengths and damping
coefficients) have been taken from \citet{2003ApJS..149..205M}. For
the oscillator strengths of Ni~II \la\la 1709, 1741, and 1751, the
values from \citet{2000ApJ...538..773F} have been adopted instead.  We
find that at least three components are required to yield an adequate
fit to the data.  Although a 3-component fit does not describe the
blue side of a few high S/N lines perfectly (see e.g. Zn~II \la 2026
in Fig.~\ref{fig:sumprofiles}), it is the simplest model that fits the
data adequately. Adding a component on the blue side would also
require an additional red component to compensate for the loss of the
broad component on the red side; the redshift of this additional red
component would be very hard to constrain. We note that the total
column density derived would hardly change if more components are
used. The redshift $z$ and the Doppler-broadening parameter $b$ (in
\kms) for each component are assumed to be the same for all ions. Only
for Mg~I we had to tweak the redshift of the red component for a
satisfactory fit. The slightly higher redshift for this component is
reasonably consistent with the profile of Mg~I \la 2852, but it is
more probable that the red Mg~I component is blended.  A third
possibility is that there are two components in this red feature,
where the reddest component would have a very high Mg~I over Zn~II
ratio.

The best-fit redshifts and broadening parameters for each component
are listed at the top of Table~\ref{tab:groundlevels}, along with the
fit ionic column densities (individual and the total of all
components). The fits are shown by the solid (red) line in
Fig.~\ref{fig:sumprofiles}.  When comparing the resonance lines with
the transitions from excited levels of Fe~II and Ni~II, it is apparent
that although the profiles are very similar, the velocity spread of
the latter is smaller. The one exception is Al~III, whose red
component (i.e. the one near +15~\kms) is not consistent with the
resonance-line fit. The column densities that we find are consistent
with those found by \citet{2006astro.ph.11092P}, with the exception of
Fe~II, where our value of log~N(Fe~II)=$15.07\pm0.08$ is lower (at
1.8$\sigma$ significance) than their adopted value of
log~N(Fe~II)=$15.22\pm0.03$. \citet{2006astro.ph.11092P} used the
transitions Fe~II \la 2249 and \la 2260, which are not saturated. In
the UVES spectra these lines fall right in the red CCD gap, leaving us
with one unsaturated line, Fe~II \la 1611, which has a lower
oscillator strength. Therefore, we have more confidence in the
determination by \citet{2006astro.ph.11092P}; in the discussion that
follows, it should be kept in mind that our total Fe~II column density
is probably too low by about 0.15~dex.

From the column densities we calculate the abundance ratios of several
ions with respect to iron for each component separately and for the
total, adopting the solar values from \citet{2003ApJ...591.1220L}. The
resulting values are listed in Table~\ref{tab:groundlevels}.  The
ratio [Zn/Fe] is high compared to the global QSO-DLA population
\citep{ledoux,2004A&A...421..479V}, and suggests a large dust
depletion, especially in component 3 where [Zn/Fe]=1.0.  The solar
value that we find for [Mn/Fe] provides additional evidence for
substantial dust depletion
\citep{2002A&A...385..802L,2006PASP..118.1077H}.  Given these
indications for a large dust depletion, the actual value for [Si/Zn]
may be 0.2-0.3 dex higher than observed ([Si/Zn]$_{\rm tot}=-0.08$),
which would suggest an $\alpha$-element overabundance, provided that
zinc can be used as a proxy for iron peak elements.  Although the
values for [Zn/Fe] and [Mn/Fe] are high compared to those found in
QSO-DLAs ([Zn/Fe]$_{\rm QSOs}$=0-1 and [Mn/Fe]$_{\rm
  QSOs}$=$-$0.5-0.4), they are rather typical for the ISM of GRB host
galaxies, with [Zn/Fe]$_{\rm GRBs}$=1-2 and [Mn/Fe]$_{\rm
  GRBs}$=0.1-0.3
\citep{2003ApJ...585..638S,2004ApJ...614..293S,2006NJPh....8..195S}. This
dust depletion difference between QSO-DLAs and GRB host galaxies can
be naturally explained if QSO sightlines on average do not probe the
central regions of galaxies \citep[for which there is growing
  evidence,
  e.g.][]{2006astro.ph..8040W,2005MNRAS.357..354E,2005ApJ...620..703C},
while GRB lines-of-sight do.

\begin{figure*}
  \centering \includegraphics[width=18cm]{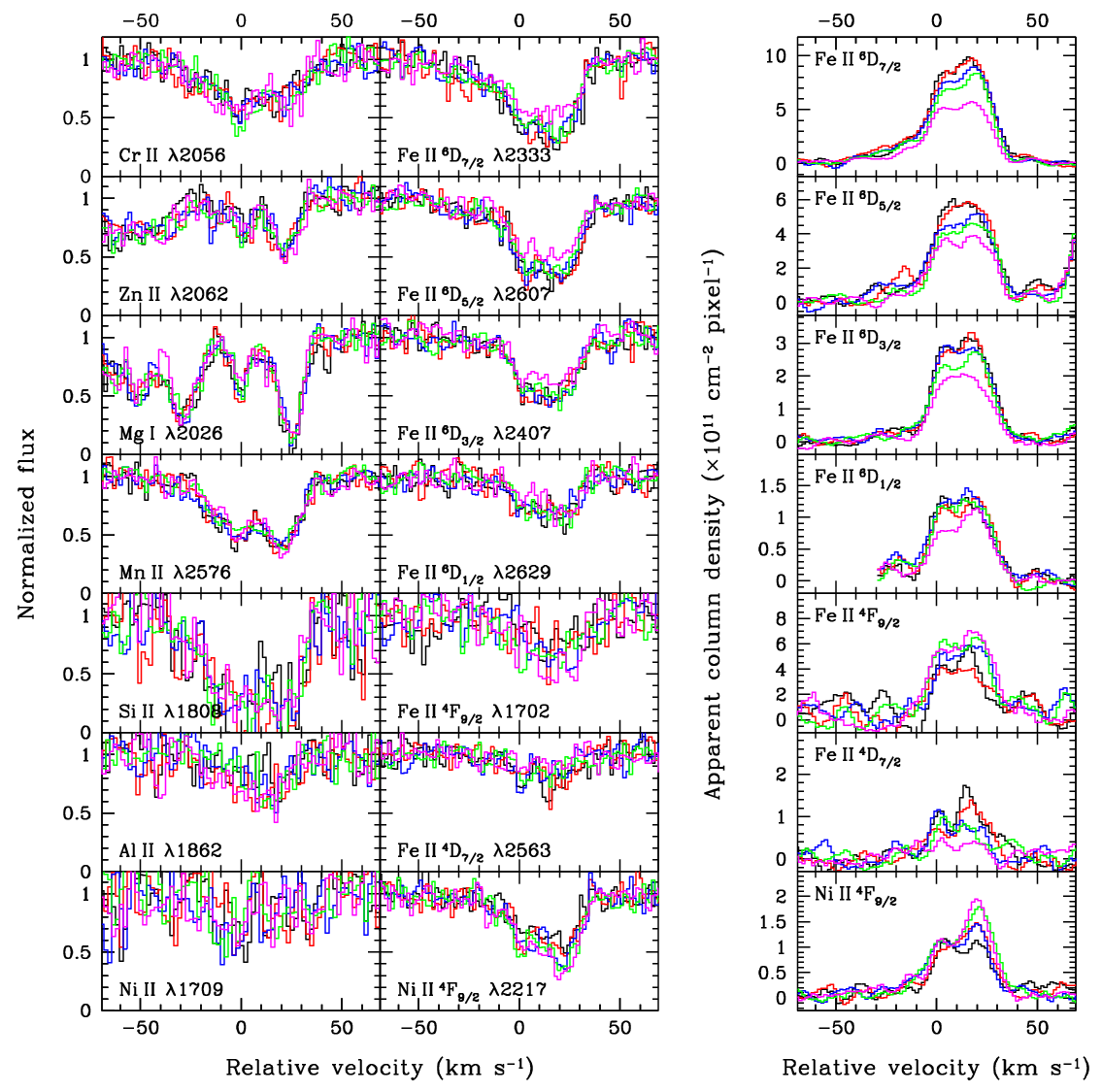}
  \caption{The epoch 1-5 UVES spectra of \grb\ (see Table
    \ref{tab:log}) are overplotted with the colours black, red, blue,
    green and magenta, respectively. In the left panel individual
    lines are shown, typical resonance lines on the left, and the
    lines arising from the excited levels of Fe~II and Ni~II on the
    right. The latter show evidence for a varying equivalent width as
    a function of time. To make this variability clearer, we have
    combined various lines that arise from the same level and
    constructed apparent column density profiles, smoothed with a
    boxcar of 5 pixels; these are shown in the right
    panel. \label{fig:profiles}}
\end{figure*}

\begin{table}[t]
  \centering 
  \caption {Column densities of the Fe~II and Ni~II excited levels, in
    the three individual components and the total, at the various
    epochs (see Table~\ref{tab:log}) since the burst
    trigger. \label{tab:metalevels}} \null\vspace{-1.0cm}
  $$
  \begin{array}{cllll}
    \hline
    \hline
    \rm epoch     & \multicolumn{4}{c}{\log~N\pm\sigma _{\log~N}$\a$} \\
    \hline
    \rm comp.     & 1            & 2            & 3            & \rm total \\
    z$\abs$       & 1.48996(6)   & 1.49002(4)   & 1.49015(9)      & \\
    b             & 28.0\pm3.5   & 5.0\pm2.1    & 10.2\pm1.5    & \\
    \hline
    & \multicolumn{4}{c}{\rm Fe~II\,^6{\rm D}_{7/2}\,$\la\la$2333, 2383, 2389, 2612, 2626}\\
    1 & 13.49\pm0.02 & 13.27\pm0.03 & 13.63\pm0.02 & 13.96\pm0.01 \\
    2 & 13.52\pm0.02 & 13.17\pm0.03 & 13.64\pm0.02 & 13.96\pm0.01 \\
    3 & 13.48\pm0.02 & 13.14\pm0.03 & 13.62\pm0.02 & 13.93\pm0.01 \\
    4 & 13.37\pm0.02 & 13.13\pm0.03 & 13.57\pm0.02 & 13.87\pm0.01 \\
    5 & 13.23\pm0.03 & 13.06\pm0.03 & 13.41\pm0.02 & 13.73\pm0.01 \\
    \hline
    & \multicolumn{4}{c}{\rm Fe~II\,^6{\rm D}_{5/2}\,$\la\la$2328, 2381, 2399, 2607, 2618}\\
    1 & 13.09\pm0.05 & 13.04\pm0.03 & 13.45\pm0.02 & 13.71\pm0.02 \\
    2 & 13.12\pm0.04 & 13.06\pm0.03 & 13.45\pm0.02 & 13.72\pm0.02 \\
    3 & 13.12\pm0.04 & 13.02\pm0.03 & 13.39\pm0.02 & 13.68\pm0.02 \\
    4 & 13.04\pm0.04 & 12.96\pm0.03 & 13.36\pm0.02 & 13.63\pm0.02 \\
    5 & 12.84\pm0.07 & 12.86\pm0.04 & 13.24\pm0.02 & 13.50\pm0.02 \\
    \hline
    & \multicolumn{4}{c}{\rm Fe~II\,^6{\rm D}_{3/2}\,$\la\la$2338, 2359, 2407, 2411, 2614, 2621} \\
    1 & 12.60\pm0.09 & 12.88\pm0.03 & 13.17\pm0.02 & 13.42\pm0.02 \\
    2 & 12.63\pm0.08 & 12.85\pm0.03 & 13.18\pm0.02 & 13.42\pm0.02 \\
    3 & 12.59\pm0.08 & 12.86\pm0.03 & 13.18\pm0.02 & 13.42\pm0.02 \\
    4 & 12.56\pm0.08 & 12.78\pm0.03 & 13.13\pm0.02 & 13.36\pm0.02 \\
    5 & 12.38\pm0.13 & 12.67\pm0.04 & 13.00\pm0.02 & 13.23\pm0.03 \\
    \hline
    & \multicolumn{4}{c}{\rm Fe~II\,^6{\rm D}_{1/2}\,$\la\la$2345, 2411, 2622, 2629} \\
    1 & 12.38\pm0.15 & 12.51\pm0.06 & 12.77\pm0.03 & 13.06\pm0.04 \\
    2 & 12.39\pm0.14 & 12.42\pm0.06 & 12.80\pm0.03 & 13.06\pm0.04 \\
    3 & 12.65\pm0.07 & 12.34\pm0.07 & 12.77\pm0.03 & 13.10\pm0.03 \\
    4 & 12.44\pm0.10 & 12.42\pm0.05 & 12.79\pm0.03 & 13.06\pm0.03 \\
    5 & 12.35\pm0.13 & 12.21\pm0.08 & 12.73\pm0.03 & 12.97\pm0.04 \\
    \hline
    & \multicolumn{4}{c}{\rm Fe~II\,^4{\rm F}_{9/2}\,$\la\la$1566$\b$, 1612$\b$, 1637$\b$, 1702$\b$, 2332, 2360} \\
    1 & 12.74\pm0.33 & 12.79\pm0.14 & 13.17\pm0.06 & 13.42\pm0.10 \\
    2 & 12.99\pm0.17 & 12.62\pm0.19 & 13.26\pm0.05 & 13.51\pm0.07 \\
    3 & 12.91\pm0.19 & 12.89\pm0.10 & 13.40\pm0.04 & 13.61\pm0.05 \\
    4 & 12.97\pm0.15 & 13.09\pm0.06 & 13.41\pm0.03 & 13.68\pm0.04 \\
    5 & 12.54\pm0.44 & 13.03\pm0.07 & 13.47\pm0.03 & 13.64\pm0.06 \\
    6 & 13.38\pm0.12 & 12.53\pm0.42 & 13.42\pm0.07 & 13.73\pm0.08 \\
    \hline
    & \multicolumn{4}{c}{\rm Fe~II\,^4{\rm D}_{7/2}\,$\la\la$1635, 2563} \\
    1 & 12.36\pm0.30 & 12.35\pm0.15 & 12.78\pm0.06 & 13.02\pm0.10 \\
    2 & 12.33\pm0.30 & 12.09\pm0.25 & 12.75\pm0.06 & 12.95\pm0.11 \\
    3 & 12.46\pm0.20 & 12.35\pm0.12 & 12.40\pm0.12 & 12.88\pm0.10 \\
    4 & 11.35\pm0.80 & 12.41\pm0.10 & 12.49\pm0.09 & 12.77\pm0.10 \\
    5 & 12.45\pm0.20 & 11.54\pm0.72 & 11.98\pm0.30 & 12.61\pm0.20 \\
    \hline
    & \multicolumn{4}{c}{\rm Ni~II\,^4{\rm F}_{9/2}\,$\la\la$2166, 2217, 2223, 2316} \\
    1 & 12.62\pm0.11 & 12.59\pm0.06 & 13.03\pm0.03 & 13.27\pm0.03 \\
    2 & 12.83\pm0.07 & 12.63\pm0.06 & 13.10\pm0.02 & 13.37\pm0.03 \\
    3 & 12.85\pm0.06 & 12.66\pm0.05 & 13.12\pm0.02 & 13.40\pm0.02 \\
    4 & 12.82\pm0.06 & 12.68\pm0.05 & 13.22\pm0.02 & 13.45\pm0.02 \\
    5 & 12.86\pm0.06 & 12.61\pm0.06 & 13.24\pm0.02 & 13.46\pm0.02 \\
    \hline
  \end{array}
  $$
  \flushleft
  \a\ The errors listed are the formal errors provided by FITLYMAN;
  although an error of 0.02 on an individual component is likely to be
  an underestimate, we consider the error on the total column density
  to be realistic.\\
  \b\ These lines are also covered in the spectrum with setting 437~nm
  (see Table~\ref{tab:log}), resulting in the determination
  of column densities at a 6$^{\rm th}$ epoch.\\
\end{table}

\begin{figure*}
  \centering
  \includegraphics[width=9.0cm,clip=true]{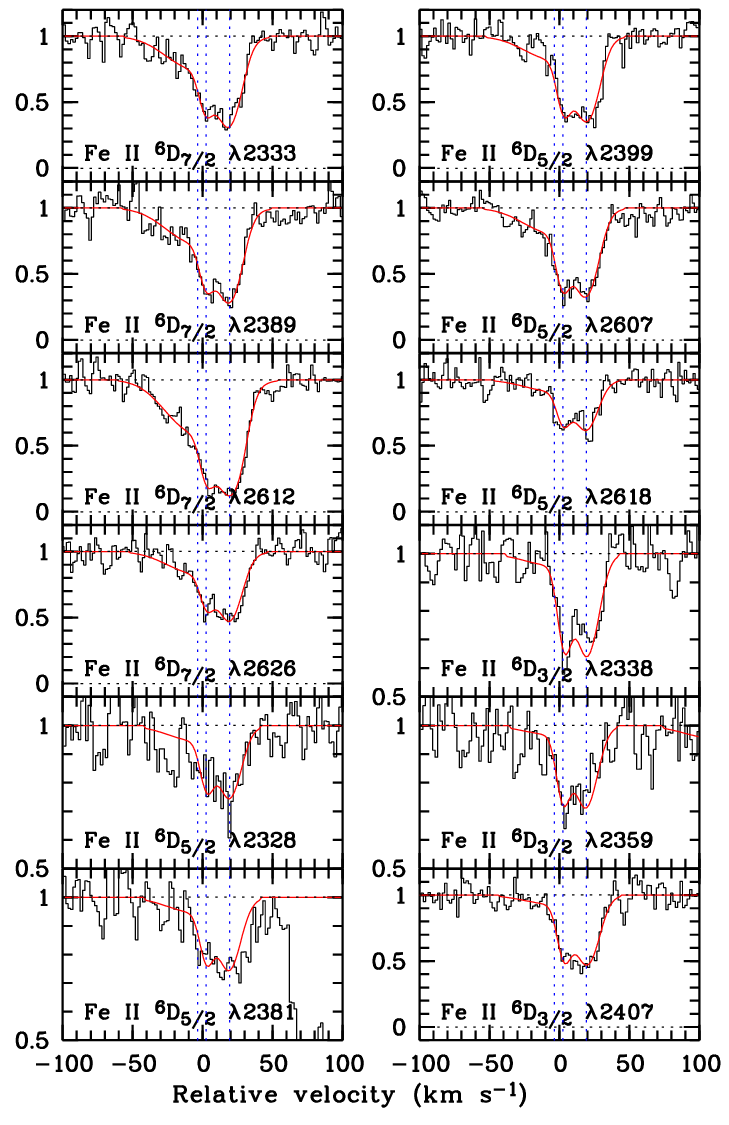}
  \includegraphics[width=9.0cm,clip=true]{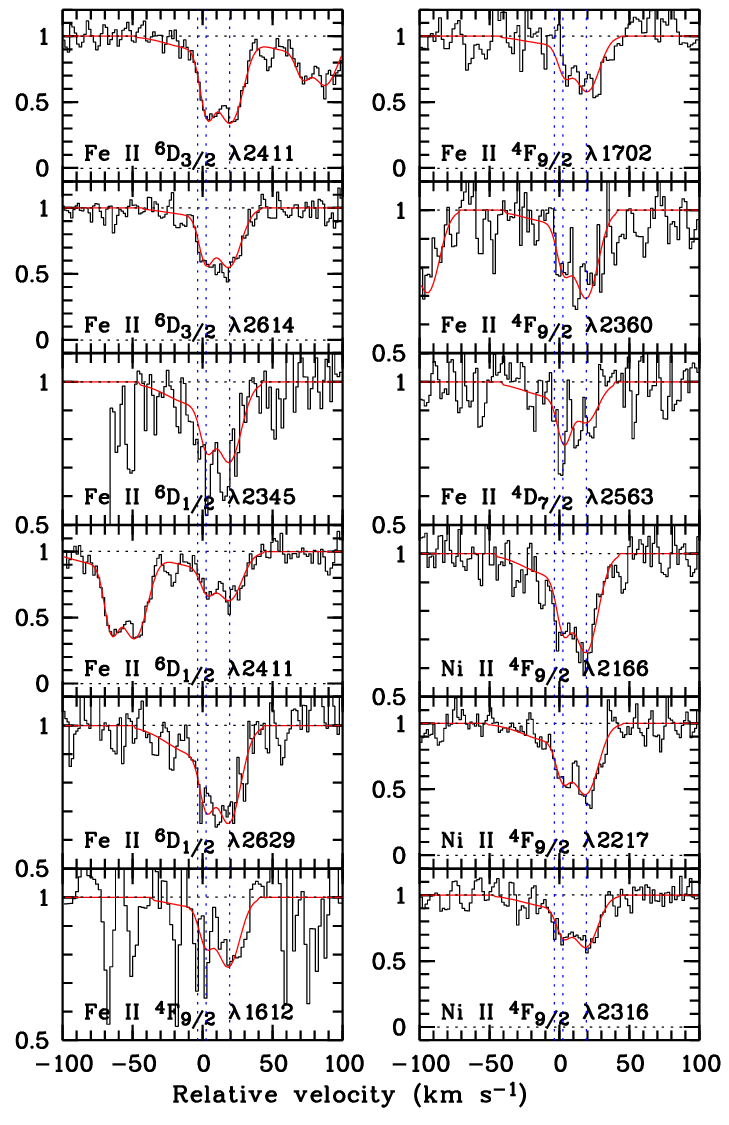}
  \caption{Absorption-line profile fits to selected transitions from
    excited Fe~II and Ni~II. The lower level of the transition, for
    which the column density is determined from the fits, is indicated
    in each panel. The fit results are listed in
    Table~\ref{tab:metalevels}.  We only show the fits for the third
    epoch spectra; the quality of the spectral fits for the other
    epochs are very similar, with only small differences due to
    slightly different signal-to-noise ratios (see
    Table~\ref{tab:log}). \label{fig:fineprofiles}}
\end{figure*}

\section{Detection and variability of transitions involving excited
  levels of Fe~II and Ni~II}
\label{sec:variability}

\begin{figure}
  \centering
  \includegraphics[width=9.0cm,clip=true]{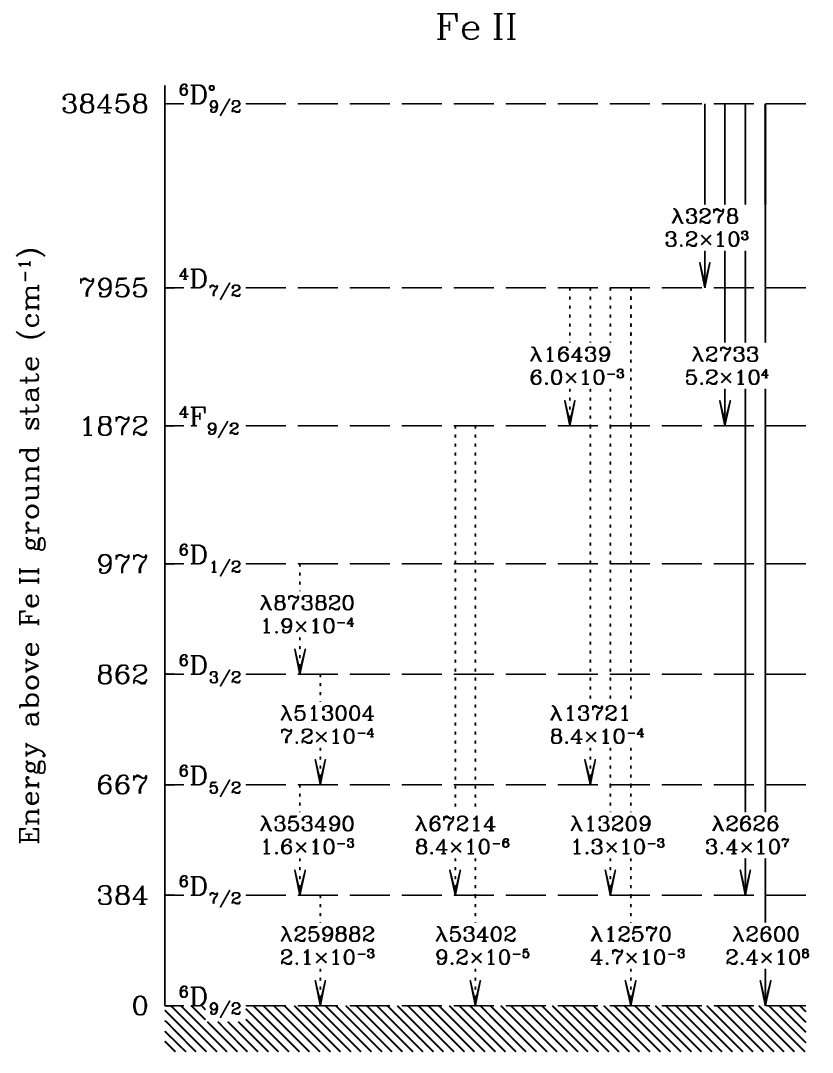}
  \caption{Energy level diagram for selected levels of Fe~II.  For the
    lower levels we only show the levels for which we detect
    transitions, i.e. the fine-structure levels of the Fe~II ground
    state, and $^4$F$_{9/2}$ and $^4$D$_{7/2}$. Note that for clarity
    reasons, we do not show the fine-structure levels of the
    latter. With the arrows, we indicate the most likely transitions
    between these levels, including one higher excited level. For each
    transition we show the wavelength in \AA\ and the spontaneous
    decay Einstein coefficient, A$_{\rm ul}$ in s$^{-1}$ (which is
    proportional to the absorption coefficient B$_{\rm lu}$, see also
    below). The electric dipole allowed transitions are indicated with
    a solid line, and the forbidden transitions (magnetic dipole or
    electric quadrupole) with a dotted line. Note that to populate the
    level $^6$D$_{1/2}$, either four IR photons are required, or two
    UV photons, where the higher levels involved need to have J=7/2
    and J=3/2. This, combined with the much larger transition
    probabilities in the UV, makes the UV pumping mechanism much more
    efficient than IR excitation.
\label{fig:FeIIatomlevels}}
\end{figure}

\begin{figure}
  \centering
  \includegraphics[width=9.0cm,clip=true]{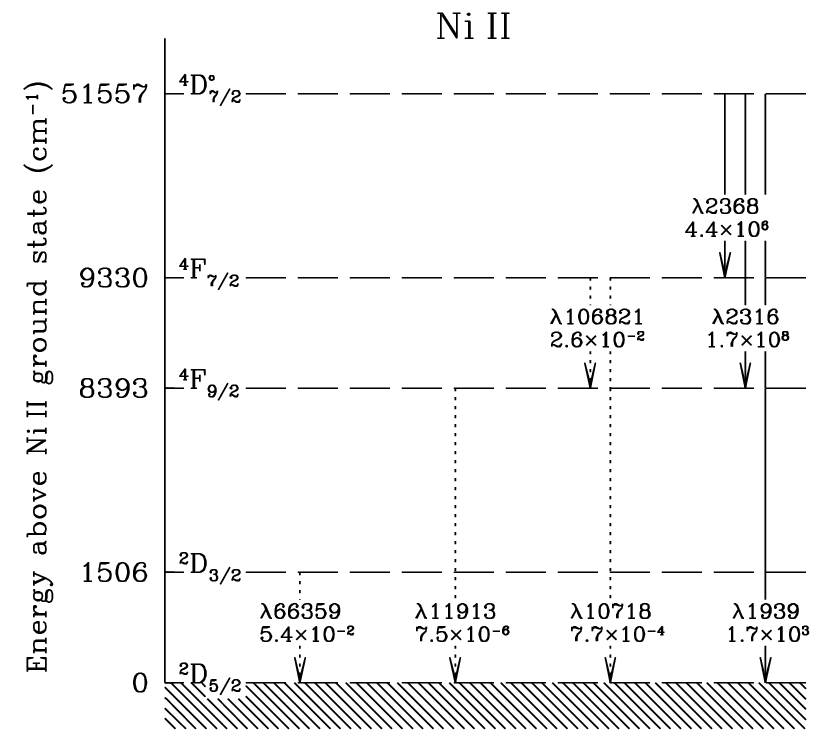}
  \caption{Energy level diagram for selected levels of Ni~II (see
    Fig.~\ref{fig:FeIIatomlevels}). Note the very low A-value for the
    transition $^4$F$_{9/2}$ to the ground state (this is actually an
    electric quadrupole transition), while the level $^4$F$_{9/2}$ can
    be very easily populated from the higher excited levels, one of
    them being $^4$D$^{\rm o}_{7/2}$. Therefore this level can be
    expected to be densely populated in the presence of a strong UV
    radiation field. The fine-structure level of Ni~II ground state,
    $^2$D$_{3/2}$ has a relatively high probability of spontaneous
    decay, with a mean lifetime of 1/(5.4$\times
    10^{-2}$~s$^{-1}$)=19~s.
    \label{fig:NiIIatomlevels}}
\end{figure}

In the left panel of Fig.~\ref{fig:profiles} we show the profiles of
some selected resonance lines and transitions arising from all four
fine-structure levels of Fe~II ($^6$D$_{7/2}$, $^6$D$_{5/2}$,
$^6$D$_{3/2}$, and $^6$D$_{1/2}$), as well as from transitions from
metastable levels of Fe~II ($^4$F$_{9/2}$ and $^4$D$_{7/2}$) and Ni~II
($^4$F$_{9/2}$). See Figs.~\ref{fig:FeIIatomlevels} and
\ref{fig:NiIIatomlevels} for an illustration of the relevant
energetically lower levels of Fe~II and Ni~II, including the first
higher excited level, and the wavelength and spontaneous decay
probability of the transitions between the levels. Back to
Fig.~\ref{fig:profiles}: we overplot the series of five spectra, epoch
1-5 (see Table~\ref{tab:log}), with the colours black, red, blue,
green and magenta, respectively. Comparison of the two panels shows
clear evidence for variability of the excited-level lines, while the
strengths of the resonance lines are constant in time. To show this
variability more clearly, we have constructed apparent column density
profiles based on pixel optical depths in composite spectra
\citep{1991ApJ...379..245S} for each of the Fe~II fine-structure
levels and the Fe~II and Ni~II metastable levels; these profiles,
which have been smoothed with a boxcar of 5 pixels, are shown in the
right panel of Fig.~\ref{fig:profiles}.

We estimate the formal significance of this variability by measuring
the equivalent width (EW) of the individual lines in between
$-30$~\kms\ and +40~\kms\ at the various epochs, conservatively adding
3\% of the EW to its formal error, due to the uncertainty in the
placement of the continuum. Using the different EW values over the
different epochs and its mean, we calculate the chi-square and the
corresponding probability with which a constant equivalent width can
be rejected. For the individual lines shown in
Fig.~\ref{fig:profiles}: Fe~II \la 2333, Fe~II \la 2607, Fe~II \la
2407, Fe~II \la 2629, Fe~II \la 1702, Fe~II \la 2563, and Ni~II \la
2217, the significances are 4.5$\sigma$, 5.8$\sigma$, 2.1$\sigma$,
0.3$\sigma$, 2.5$\sigma$, 1.7$\sigma$, and 3.5$\sigma$, respectively.
Using several transitions originating from the same level (the same
that have been used to construct the apparent column density profiles
in Fig.~\ref{fig:profiles}), we find the following numbers:
8.7$\sigma$ (Fe~II $^6$D$_{7/2}$), 7.4$\sigma$ (Fe~II $^6$D$_{5/2}$),
3.2$\sigma$ (Fe~II $^6$D$_{3/2}$), 0.5$\sigma$ (Fe~II $^6$D$_{1/2}$),
2.2$\sigma$ (Fe~II $^4$F$_{9/2}$), 1.7$\sigma$ (Fe~II $^4$D$_{7/2}$),
and 2.5$\sigma$ (Ni~II $^4$F$_{9/2}$).

We have performed Voigt-profile fitting to the lines originating from
the excited levels of Fe~II and Ni~II, independent from the
resonance-line fit. The atomic data from \citet{2003ApJS..149..205M}
were adopted when available, and if not we have assumed the values
from \citet{2003IAUS..210...45K} (note that for Ni~II we have divided
the Kurucz oscillator strengths by two; see the discussion in
Sect.~\ref{sec:radiation}). The Voigt-profile fit results are shown in
Table~\ref{tab:metalevels} and Fig.~\ref{fig:fineprofiles}; we only
show the fit profiles for epoch 3 as the other epochs display very
similar results, but with somewhat different signal-to-noise ratios.
Just as with the resonance lines, a satisfactory fit is found when
using three components. In an initial fit, the redshift and $b$
parameter were free to vary from epoch to epoch. As these were found
to be constant with time, in a final fit they were fixed to the
averages over the five epochs; the $z$ and $b$ averages and the
corresponding standard deviations are listed at the top of
Table~\ref{tab:metalevels}. The column density errors listed in
Table~\ref{tab:metalevels} are the formal errors provided by FITLYMAN.
We have also estimated an error in the placement of the continuum by
varying the sigma clipping factors and the order of the polynomial
with which we fit the continuum around each line, and rerun the Voigt
profile fit. The maximum change that we find is 0.03 dex, which we add
to the formal error for the rest of the analysis.

\citet{2006astro.ph.11092P} have performed time-resolved
high-resolution spectroscopy of \grb\ as well. Their three spectra
were taken around the same mid-exposure time as our epoch 4, 5 and 6
spectra. They do not consider variability, and report on the average
column density for the four Fe~II fine-structure levels; they do not
mention the metastable levels of Fe~II and Ni~II. When comparing their
average values with our epoch 4 column densities, the results are
fully consistent within the errors.

Comparison of the resonance-lines fit with the excited-lines fit shows
that the redshifts and $b$ parameters for the three components are
very similar, cf. Tables~\ref{tab:groundlevels} and
\ref{tab:metalevels}. When we run the excited-lines profile fit with
the redshift and $b$ parameter fixed at the values of the resonance
lines, the resulting fit is very poor. Therefore, although the fits
provide similar results, the redshifts and $b$ parameters are
significantly different. We already noted this difference in
Sect.~\ref{sec:hostabsorbers}, and we will come back to this point in
Sect.~\ref{sec:discussion}.

Atomic fine-structure levels are caused by an energy split due to the
interaction of the total electron spin and total angular momentum of
the electrons. The transitions between these levels are not allowed,
i.e. they cannot proceed through an electric dipole transition, and
therefore the corresponding transition probabilities are low. The same
is applicable to other energetically lower levels, also called
metastable levels, and their fine-structure
levels. Figures~\ref{fig:FeIIatomlevels} and \ref{fig:NiIIatomlevels}
show the energy level diagrams, of selected levels of Fe~II and Ni~II.

These fine-structure and metastable levels can be populated through
(1) collisions between the ion and other particles such as free
electrons, (2) direct photo-excitation by infra-red (IR) photons (with
specific wavelengths between 87-260~$\mu$m), and/or (3) indirectly
through excitation by ultra-violet (UV) photons, followed by
fluorescence. Detection of transitions from these energetically lower
excited levels provides a powerful probe of the physical conditions in
the interstellar medium \citep{1968ApJ...152..701B}, where the
quantities that can be derived depend on the excitation mechanism.

\citet{vrees030323} noted the presence of transitions originating from
the fine-structure level of Si~II in the host galaxy of GRB~030323; as
these lines had never been clearly detected in QSO-DLAs \citep[we note
  that they had been detected in absorption systems associated with
  the QSO, see][]{1995ApJ...443..586W,2001A&A...373..816S}, this
detection suggested an origin in the vicinity of the GRB. Assuming
that collisions with electrons were the dominant excitation mechanism,
a volume density of n$_{\rm HI}=10^2-10^4$~cm$^{-3}$ was derived
\citep[see also][]{2004ApJ...614..293S,2006A&A...451L..47F}. We note
that Si~II $^2$P$_{3/2}$ \la 1816 is not detected (5~$\sigma$ upper
limit: log N$<$14.81) in the case of \grb, and that Si~II
$^2$P$_{3/2}$ \la 1533 (see Fig.~\ref{fig:sumprofiles}) is severely
blended with Fe~II \la 2382 at the redshift of an intervening absorber,
$z=0.603$. Therefore, this (or any) Si~II fine-structure level is not
included in our analysis.

More recently, even more exotic transitions involving fine-structure
levels of Fe~II have been discovered in GRB sightlines
\citep{2005ApJ...634L..25C,2006ApJ...646..358P,2006astro.ph..1057P,2006astro.ph..9825D}. As
noted by \citet{2006astro.ph..1057P}, these lines had previously been
detected in absorption in extreme environments such as Broad
Absorption-Line (BAL) quasars \citep{2002ApJS..141..267H}, $\eta$
Carinae \citep{2005ApJ...620..442G}, and the disk of $\beta$ Pictoris
\citep{1988A&A...190..275L}. For the Fe~II fine-structure level
population along GRB sightlines it has been argued
\citep{2006astro.ph..1057P} that IR excitation is negligible, that
collisional excitation is improbable (although not excluded), and that
indirect UV pumping probably is the dominant excitation mechanism. The
detection of variability at the 3$\sigma$ level (using two different
instruments) of one Fe~II fine-structure line in the spectrum of
GRB~020813 was reported \citep{2006astro.ph..6462D}, which the authors
claim to be supportive evidence for the UV pumping model.

The lines arising from the metastable levels of both Fe~II
($^4$F$_{9/2}$ and $^4$D$_{7/2}$) and Ni~II ($^4$F$_{9/2}$) that we
detect are the first lines from metastable levels to be identified
along any GRB sightline. However, these have also been previously
detected in BAL quasars
\citep{1987ApJ...323..263H,1995ApJ...443..586W} and $\eta$ Carinae
\citep{2005ApJ...620..442G}.

Our clear detection of time-variation of numerous transitions
involving all fine-structure levels of the Fe~II ground state, and
moreover transitions originating from metastable levels of Fe~II and
Ni~II, allows for a critical comparison of the data with the three
possible excitation mechanisms mentioned above.  However, independent
of the mechanism at play, the detection of time-variable absorption
implies that the flux from the GRB prompt emission and/or afterglow,
directly or indirectly, is the cause of the line variability, and that
the absorbing atoms are located in the relative vicinity of the GRB
explosion.

\section{Modeling of the time evolution of Fe~II and Ni~II excited
levels}
\label{sec:modeling}

\subsection{Collisional excitation}
\label{sec:collisions}

We first consider the collisional model.  Although
\citet{2006astro.ph..1057P} did not detect any change in column
density of excited Fe~II, they did suggest that variability of
absorption-line strengths would be inconsistent with a collisional
origin of the excitation. This is certainly expected to be the case
for a medium out of reach of the influence of the GRB afterglow flux,
but close to the GRB site one might expect the incidence of intense
X-ray and UV radiation to deposit a considerable amount of energy in
the surrounding medium through photo-ionization, causing a situation
similar to photo-dissociation regions (PDRs), which can produce a
shock front with typical velocities of 10-20~\kms\ and density
enhancements. Relaxation of these high-density regions might then
result in a change in column density.  The profiles shown in
Fig.~\ref{fig:sumprofiles} are actually suggestive of such a shock
front situation: they show two main components with a velocity
difference of 25~\kms, and moreover, the lines caused by transitions
from the metastable levels seem to be enclosed by the ground state
species of e.g. Cr~II. Therefore, it is quite reasonable to consider
the collisional excitation scenario.

\begin{figure}[bht]
  \centering \includegraphics[width=9cm]{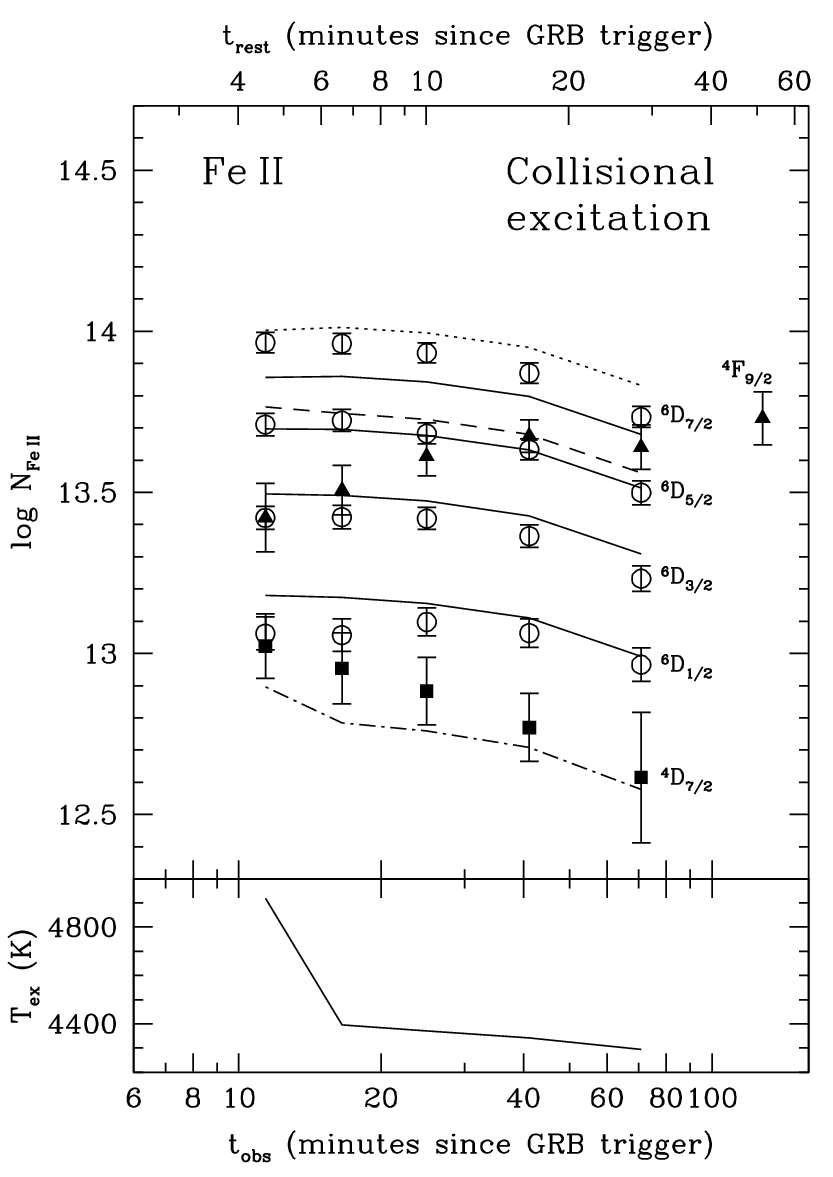}
  \caption{The top panel shows the observed total column densities
    (see the last column of Table~\ref{tab:metalevels}) for the
    fine-structure lines (open circles; from top to bottom:
    $^6$D$_{7/2}$, $^6$D$_{5/2}$, $^6$D$_{3/2}$, and $^6$D$_{1/2}$,
    respectively), the first metastable level (filled triangles,
    $^4$F$_{9/2}$), and the second metastable level (filled squares,
    $^4$D$_{7/2}$) of Fe~II. Overplotted are the results of the
    best-fit Boltzmann model (collisional scenario): solid lines for
    the fine-structure levels, dashed line for $^4$F$_{9/2}$, and
    dashed-dotted for $^4$D$_{7/2}$. The best-fit Fe~II ground state
    column density is shown by the dotted line, while the best-fit
    excitation temperature (T$_{\rm ex}$) is depicted in the bottom
    panel. It is clear that the Boltzmann model can be rejected with
    high confidence. \label{fig:boltzmann}}
\end{figure}

If collisions of the Fe II ions with electrons, protons or HI atoms is
the dominant excitation process, and the collisional de-excitation
rate exceeds the spontaneous decay rate, the population ratio between
two levels $i$ and $j$ should follow the Boltzmann distribution
\citep[see e.g.][]{2006astro.ph..1057P}:
\begin{equation}
  \frac{n_i}{n_j} = \frac{g_j}{g_i} e^{-E_{ij}/k T_{ex}}
  \label{eq:boltzmann}
\end{equation}
where g is the statistical weight of the level, $E_{ij}$ is the energy
jump between the levels, $k$ is the Boltzmann constant, and $T_{ex}$
is the excitation temperature. We fit this Boltzmann distribution to
all available excited levels of Fe~II at each epoch separately. For
each epoch we fit two parameters: the density (in our case column
density) and the excitation temperature. The resulting Boltzmann fit
is shown in Fig.~\ref{fig:boltzmann}. Even with 2 free parameters for
each of the five epochs, the model is clearly not able to reproduce
the observed column densities: the reduced chi-square is
$\chi^2_{\nu}=95.7/(5-2)=31.9$.

If we would have only detected the fine-structure levels of Fe~II, the
collisional excitation model fit would have been acceptable, as in
\citet{2006astro.ph..1057P}. The main culprit for the poor fit is the
increasing level population of the metastable level Fe~II
$^4$F$_{9/2}$, which cannot be accommodated in the Boltzmann fit
because all the other levels, with similar energies, are decreasing
with time.  Inclusion of Ni~II $^4$F$_{9/2}$ (E=8393~cm$^{-1}$) would
make the fit even worse, as its energy level is similar to the Fe~II
$^4$D$_{7/2}$ level (E=7955~cm$^{-1}$), while its observed column
density is increasing with time (see the bottom panel of
Fig.~\ref{fig:uv}). Moreover, the best-fit Boltzmann model predicts
column densities for the fine-structure levels of Fe~II $^4$F$_{9/2}$
(e.g. the predicted column densities for its first fine-structure
level, Fe~II $^4$F$_{7/2}$, is log~$N=13.4-13.6$) that are
inconsistent with the upper limits we obtain for this level (see
Fig.~\ref{fig:uvlimits}). To preserve clarity, we do not show these in
Fig.~\ref{fig:boltzmann}.

An implicit assumption in this collisional excitation model is that of
local thermodynamic equilibrium (LTE), while the observed variability
of the absorption lines suggests that this may not be valid. However,
using the PopRatio\footnote{We note that PopRatio only includes
  collisions with electrons, which are the dominant collision partners
  for temperatures below approximately 100,000~K; beyond this
  temperature the contribution of collisions with protons and HI atoms
  become significant \citep[see Figs.~3 and 7
    of][]{2002MNRAS.329..135S}} code \citep{2002MNRAS.329..135S}, we
find that if collisions is the dominant excitation mechanism, the
observed population ratios of the Fe~II fine-structure levels require
an electron volume density of at least $n_e\sim 10^4$~cm$^{-3}$.  As
this is very high, while at the same time the observed variability is
relatively smooth in time, the assumption of LTE probably is valid.

In conclusion, the collisional model is rejected with high confidence.

\subsection{Radiative excitation by GRB-afterglow photons}
\label{sec:radiation}

To verify if our observations can be explained by radiative excitation
(by IR and/or UV photons), we now consider a model of a cloud with
column density $N$ (atoms cm$^{-2}$), at a distance $d$ (pc) from the
GRB. The afterglow flux will excite the atoms in the cloud, and we
will calculate the atom level populations as a function of time, to be
compared with our observations. We will only consider excitation, and
neglect ionization, which, as we will see below, is fully justified.

We can describe the afterglow flux in the host-galaxy rest frame by:
\begin{equation}
  F^{\rm rest}_{\nu} = \frac{
    1.192\times10^{-25} 
    \left[\frac{t_{\rm obs}}{393~{\rm s}}\right]^{\alpha} 
    \left[\frac{\lambda_{\rm obs}}{5439~{\rm\AA}}\right]^{-\beta}
    \left[\frac{1.083\times10^{10}~{\rm pc}}{d}\right]^2}
  {1+z}
  \label{eq:fnu}
\end{equation}
in erg s$^{-1}$ cm$^{-2}$ Hz$^{-1}$, where we have used the V=14.99
UVOT measurement at 393 seconds after the burst
\citep{2006GCN..4978....1S}, corrected for foreground extinction in
the Galaxy with $A_{\rm V}$=0.74 \citep{1998ApJ...500..525S}, and in
the absorber at $z=1.1$ (which shows a clear 2175~\AA\ extinction
bump) with a Milky Way extinction curve and $A_{\rm V}=0.25$ (at
$z=1.1$), resulting in an effective $A_{\rm V, z=0}=0.55$
\citep{sara060418}. So the constant in Eq.~\ref{eq:fnu} is the
would-be observed UVOT flux at $z=0$ if there would not have been any
foreground extinction. The best-fit afterglow intrinsic spectral slope
assuming these extinction values is $\beta=-0.8$
\citep[see][]{sara060418}. However, as this value is quite uncertain
we will also determine a best-fit value for $\beta$ in the fit. The
flux decay in time of our spectra is very well described by a power
law with index $-1.1$, which we adopt for $\alpha$. The value for the
decay index determined from UBVRIz photometry data reported in GCNs
range from $-1.1$ to $-1.3$
\citep{Nysewander06_GCN4971,Cobb06_GCN4972,2006GCN..4978....1S}.  For
the calculation of the luminosity distance to the GRB,
d$_l=1.083\times10^{10}$~pc, we have adopted
H$_0=70$~km~s$^{-1}$~Mpc$^{-1}$, $\Omega_{\rm M}=0.3$, and
$\Omega_{\Lambda}=0.7$.

The atom level population of an upper level $u$ with respect to a
lower level $l$ is given by the balance equation (but see erratum in
Appendix A for the correct version):
\begin{equation}
  \frac{dN_{\rm u}}{dt} = N_{\rm l} B_{\rm lu} F_{\nu}(\tau_0) -
  N_{\rm u} \left[A_{\rm ul}+B_{\rm ul} F_{\nu}(\tau_0)\right]
  \label{eq:dndt}
\end{equation}
where A$_{\rm ul}$, B$_{\rm ul}$ and B$_{\rm lu}$ are the Einstein
coefficients for spontaneous decay, stimulated emission and
absorption, respectively, with B$_{\rm ul}$ = A$_{\rm ul} \lambda^3 /
2 h c$ (all in cgs units), and B$_{\rm lu}$ = B$_{\rm ul}$ g$_{\rm u}$
/ g$_{\rm l}$ (g is the statistical weight of the energy level, with
g=2J+1, and where J is the total angular momentum of the electrons).
$F_{\nu}(\tau_0)$ is the incoming afterglow flux at the monochromatic
frequency corresponding to the transition energy, and modified by the
optical depth at line center \citep[see Eq.~3.8
  of][]{2005ism..book.....L}:
\begin{equation}
  F_{\nu}(\tau_0) = F_{\nu}(0) e^{-\tau_0} + S_{\nu}(1-e^{-\tau_0})
  \label{eq:fnutau}
\end{equation}
with $\tau_0 = \frac{1.497\times10^{-2} N_{\rm l} \lambda}{b} f$, and
where the oscillator strength $f$ is calculated from A$_{\rm ul}$, 
using:
\begin{equation}
  f = \frac{m_{\rm e} c A_{\rm ul} g_{\rm u} \lambda^2}{8 \pi^2
    q_{\rm e}^2 g_{\rm l}}
  \label{eq:A2f}
\end{equation}
$S_{\nu}$ is the source function \citep[see Eq.~3.6
  of][]{2005ism..book.....L}.  The Doppler width, or broadening
parameter, $b$, has been determined from the line-profile fits to be
28~\kms, 5~\kms, and 10~\kms\ for the three different components (see
Table~\ref{tab:metalevels}). We allow the $b$ value to vary with the
aim to obtain the best-fit value, to be compared with the above
measurements. For many UV transitions the cloud that we model will be
optically thick, and therefore we slice up the cloud in a sufficient
number of plane-parallel layers, so that each layer can be considered
optically thin for a particular transition; we set the maximum allowed
optical depth for a layer to be $\tau_{\rm layer max}$=0.05.

An important ingredient in the model fit is the adopted atomic data
values for the spontaneous decay coefficients A$_{\rm ul}$ (or
equivalently, $f$, see Eq.~\ref{eq:A2f}), and that these are exactly
the same as used in the Voigt profile fits performed to obtain the
observed column densities (see Tables~\ref{tab:groundlevels} and
\ref{tab:metalevels}). We made sure that this is the case. For Fe~II
we include the 20 lower energy levels in our calculations (up to
E=18886.78~cm$^{-1}$), and the A's between all these lower levels are
taken from \citet{1996A&AS..120..361Q}. For the transitions between
the lower and higher excited levels, we adopt the values by
\citet{2003ApJS..149..205M} if available, and if not then we use those
provided by \citet{2003IAUS..210...45K}\footnote{see
  http://kurucz.harvard.edu}. The number of Fe~II higher excited
levels included is 456, with a resulting number of transitions of
4443.  For Ni~II we include the lower 17 energy levels, and take the
A's between these from \citet{1996A&AS..119...99Q}, complemented by
those from \citet{1982A&A...110..295N}.  For the Ni~II ground state
transitions corresponding to Ni~II \la 1317 and 1370 we adopt the $f$
values of \citet{2006ApJ...637..548J}, and for Ni~II \la\la 1454,
1709, 1741 and 1751 from \citet{2000ApJ...538..773F}. For the other
transition probabilities between the lower and higher excited levels
of Ni~II we again use the value from \citet{2003ApJS..149..205M} if
available, and otherwise those from \citet{2003IAUS..210...45K}. As
the ratio of the $f$-values of the Jenkins et al. and Fedchak et
al. ground-state lines compared to the Kurucz values varies from 1.87
to 2.59, and we find similar ratios between values of two Morton Ni~II
fine-structure lines and those of Kurucz, we have divided all Kurucz
Ni~II A's by a factor of two. We stress that although this factor of
two results in different inferred column densities, it does not affect
the fit results since we use exactly the same oscillator strengths in
our model. The number of Ni~II higher excited levels included is 334,
with a resulting number of transitions of 3136.

We have written an IDL routine that incorporates the equations above
and the adopted Fe~II and Ni~II atomic data values, and calculates the
level evolution of the atoms in the cloud as a function of time. This
model is fit to the observations (using Craig Markwardt's MPFIT
routines\footnote{see
  http://cow.physics.wisc.edu/\~{}craigm/idl/idl.html}) with the
following free parameters: the distance $d$, the total Fe~II or Ni~II
column density $N$, the afterglow spectral slope $\beta$, the Doppler
parameter $b$, and the rest-frame time at which we start the
calculations, $t_0$. We note that this $t_0$ does not provide any
constraints on the shape of the light curve before the time that our
first spectrum was taken (we simply extrapolate the light curve back
to $t_0$ assuming a decay index of $\alpha=-1.1$), but it does
constrain the total number of photons that arrived at the cloud since
the GRB trigger.

By selecting the levels that we loop through, we can either treat IR
excitation and UV pumping separately, or combine the two in a
consistent manner. For the UV transitions, we assume that the higher
excited levels are merely a route to any lower level that the higher
level can combine with, i.e.  after excitation of a number of atoms in
one timestep, all electrons are re-distributed immediately among all
possible lower levels and no electrons stay in the higher excited
level. This is justified by the very large spontaneous decay
transition probabilities for nearly all higher excited energy
levels. As a consistency check, we compared the results of our program
with the PopRatio code \citep{2002MNRAS.329..135S}, which computes the
Fe~II fine-structure level population assuming an equilibrium
situation, i.e. $\frac{dN_{\rm u}}{dt}=0$ in Eq.~\ref{eq:dndt}. Using
exactly the same Galactic UV background as in PopRatio (converted to
flux density by applying a factor of 4~$\pi$), and in the optically
thin limit, the results are identical down to the 0.07\% level.

\begin{figure}[htb]
  \centering \includegraphics[width=9cm]{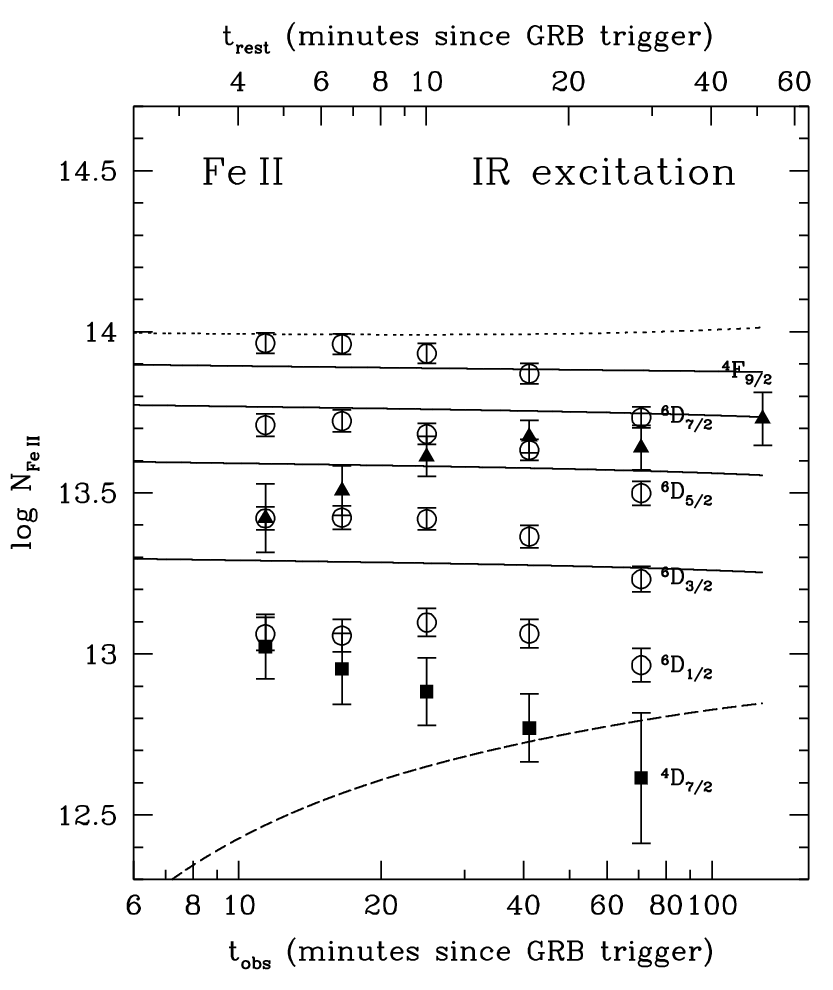}
  \caption{The top panel shows the same as the top panel of
    Fig.~\ref{fig:boltzmann}, but now with the IR excitation model
    overplotted: solid lines for the fine-structure levels, dashed
    line for $^4$F$_{9/2}$. The $^4$D$_{7/2}$ level fit column density
    does not even reach the lower limit of the plotting range. The
    model prediction for the evolution of the Fe~II ground state
    column density is shown by the dotted line. \label{fig:ir}}
\end{figure}

\subsubsection{IR excitation only}
\label{sec:ir}

First we consider only IR photons to be exciting the atoms in the
cloud. The 20 lower energy levels of Fe~II are included, and we do not
consider Ni~II. The resulting fit is shown in Fig.~\ref{fig:ir}.  We
note that we have imposed a lower limit to the distance of 2~pc and we
fixed the spectral slope at $\beta=-0.8$ and the $b$ parameter at
18~\kms; the value for the latter is unimportant as all IR transitions
are basically optically thin. For distances lower than 2~pc, the
calculation would take too long to compute on our workstation. The
reason for this is that we adjust the program timestep in such a way
that a maximum of 5\% of all atoms can be excited to the higher
excited level of a particular transition in each timestep; for a large
photon flux this requires a very small timestep, i.e. a large number
of calculations. If we would allow the distance to go under 2~pc, the
fitting routine would try to move the levels $^4$F$_{9/2}$ (dashed
line) and $^4$D$_{7/2}$ (below the lower limit of the plotting range,
peaking at log~$N = 11.6$) up, which would also cause the
fine-structure levels (solid lines) to move up slightly.  The final
chi-square would be lower than for the present $\geq$ 2~pc fit, which
has $\chi^2_{\nu}({\rm IR})= 2571/(31-3) = 91.8$, but it would still
provide an extremely poor fit to the observations. Moreover, at such a
short distance, most of the Fe~II would be expected to be ionized in
the first place
\citep[e.g.][]{2000ApJ...537..796W,2002ApJ...580..261P}, and therefore
we can safely reject IR excitation mechanism as the explanation for the
observed level population and evolution.

The reason for the relatively low population of the metastable
levels compared to the Fe~II ground state fine-structure levels in the
IR excitation case is not due to a lower transition probability for
the former; e.g. between the ground state and its first fine-structure
level $^6$D$_{7/2}$, A=2.13$\times 10^{-3}$~s$^{-1}$, while between
the ground state and the second metastable level $^4$D$_{7/2}$,
A=4.74$\times 10^{-3}$~s$^{-1}$ (see
Fig.~\ref{fig:FeIIatomlevels}). The reason is the wavelength
dependence to the third power of the Einstein absorption coefficient
B$_{\rm lu}$ (see below Eq.~\ref{eq:dndt}): photons with a longer
wavelength are much more likely to be absorbed. For the levels
mentioned above, this makes the transition from the ground state to
$^4$D$_{7/2}$ a factor of (7955/385)$^3$ $\sim$ 9000 less likely,
while the difference in the observed column density is only a factor
of 10. Had we only observed the variation of the fine-structure levels
of the ground state, and not the levels $^4$F$_{9/2}$ and
$^4$D$_{7/2}$, we would have not been able to reject the IR excitation
model with such high confidence, as merely considering those levels
results in an excellent fit to the data, with $\chi^2_{\nu}({\rm
  IR_{\rm 5 levels}})= 11.0/(20-3) =
0.65$. \citet{2006astro.ph..1057P} rejected the IR excitation scenario
on the basis that IR pumping is negligible at the distance limit set
by the detection of Mg~I in their spectra (which assumes that Mg~I and
the excited material is at the same location, which need not be the
case; see also Sect.~\ref{sec:discussion}), combined with the
observation that UV pumping is dominant at any given distance from the
GRB, in the absence of severe extinction. Although these arguments are
strong, they are not as conclusive as our modeling results.

\begin{figure}
  \centering \includegraphics[width=9cm]{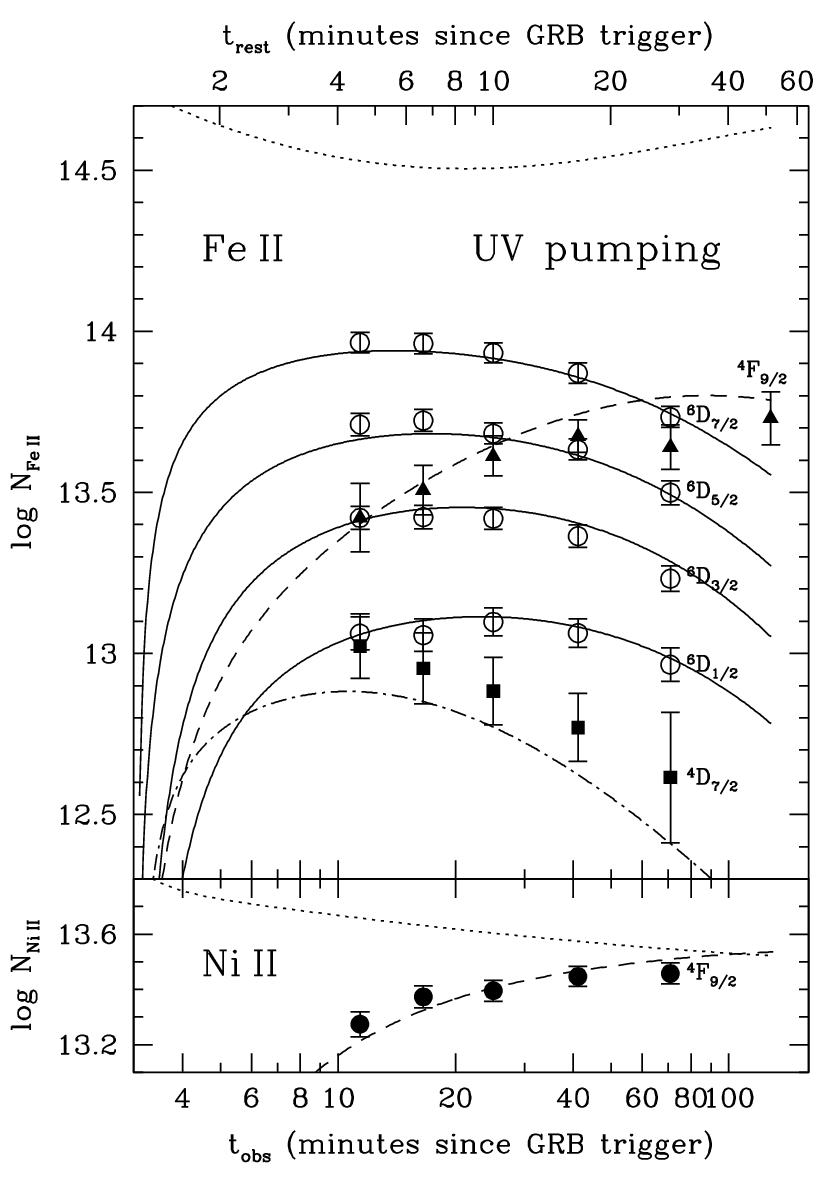}
  \caption{The top panel shows the same as the top panel of
    Fig.~\ref{fig:boltzmann}, but now with the UV pumping model
    overplotted: solid lines for the fine-structure levels, dashed
    line for $^4$F$_{9/2}$, and dashed-dotted for $^4$D$_{7/2}$.  The
    bottom panel displays the observed total column densities
    for Ni~II $^4$F$_{9/2}$ (filled circles), and the best-fit Ni~II
    model. In this Ni~II fit, all parameters except for Ni~II column
    density were fixed to the best-fit values obtained from the Fe~II
    fit. The model prediction for the evolution of the Ni~II ground
    state column density is shown by the dotted line. All Fe~II and
    Ni~II column densities are very well described by the UV pumping
    model. \label{fig:uv}}
\end{figure}

\subsubsection{UV pumping}
\label{sec:uv}

After rejection of collisional and IR excitation, we now consider the
UV pumping scenario. In the UV model calculations we consider 20 lower
and 456 higher excited levels of Fe~II. The resulting fit is shown in
the top panel of Fig.~\ref{fig:uv}. The best-fit values for the fit
parameters are as follows: log~$N$(Fe~II ground
state)=\gpm{14.75}{0.06}{0.04}, $d=1.7\pm0.2$~kpc (but see erratum in
Appendix A), $\beta=$\gpm{-0.5}{0.8}{1.0}, t$_0$=\gpm{74}{12}{11}~s,
and $b=25\pm3$~\kms, and a chi-square of $\chi^2_{\nu}({\rm
  UV-Fe~II})=26.2/(31-5)=1.01$. Next, we also model the evolution of
the Ni~II $^4$F$_{9/2}$ level, using 17 lower and 334 higher levels of
Ni~II. We fix all parameters in the Ni~II fit to the best-fit values
of the Fe~II fit, except for the Ni~II ground state column
density. The resulting fit is shown in the bottom panel of
Fig.~\ref{fig:uv}. The reduced chi-square is $\chi^2_{\nu}({\rm
  UV-Ni~II})=5.6/(5-1)=1.4$, and the best-fit Ni~II column density is
log~$N$(Ni~II ground state)=$13.84\pm0.02$. When also including the
distance as a free parameter, we find log~$N$(Ni~II ground
state)=$13.73\pm0.02$, and $d=1.0\pm0.3$~kpc (but see Appendix A),
with a chi-square of $\chi^2_{\nu}({\rm UV-Ni~II})=0.72/(5-2)=0.24$.

From Figs.~\ref{fig:FeIIatomlevels} and \ref{fig:NiIIatomlevels}, it
is straightforward to see why the levels Fe~II $^4$F$_{9/2}$ and Ni~II
$^4$F$_{9/2}$ increase with time in the UV pumping scenario. The route
to these levels is rather quick: one out of every 5000 photons at
2600~\AA\ will bring the ion to this excited level. We note that the
higher excited level shown in Fig.~\ref{fig:FeIIatomlevels} is just
one out of many levels that allow population of the Fe~II
$^4$F$_{9/2}$ level through absorption of a UV photon, followed by
spontaneous decay. Once in this level, it takes
1/(9.2$\times10^{-5}$~s$^{-1}$) = 3.0~hours for the ion to decay to
the Fe~II ground state; this is longer than the time scale over which
our spectra were recorded (1~hour in the rest frame), and explains why
this level continues to rise in Fig.~\ref{fig:uv}.  Ni~II
$^4$F$_{9/2}$ is even easier to populate through the absorption of UV
photons, and will take a longer time to decay to the Ni~II ground
state: 37~hours. Transitions arising from these Fe~II and Ni~II
metastable levels are therefore excellent probes of the UV pumping
mechanism, as they can be observed up to many hours after the GRB
event. In fact, although we were the first to identify them, these
lines should also be present in the high-resolution spectra of
GRBs~050730, 051111\citep{2006astro.ph..1057P,2006astro.ph.11092P} and
050922C \citep{D'Elia05_GCN4044}.

\begin{figure}
  \centering \includegraphics[width=9cm]{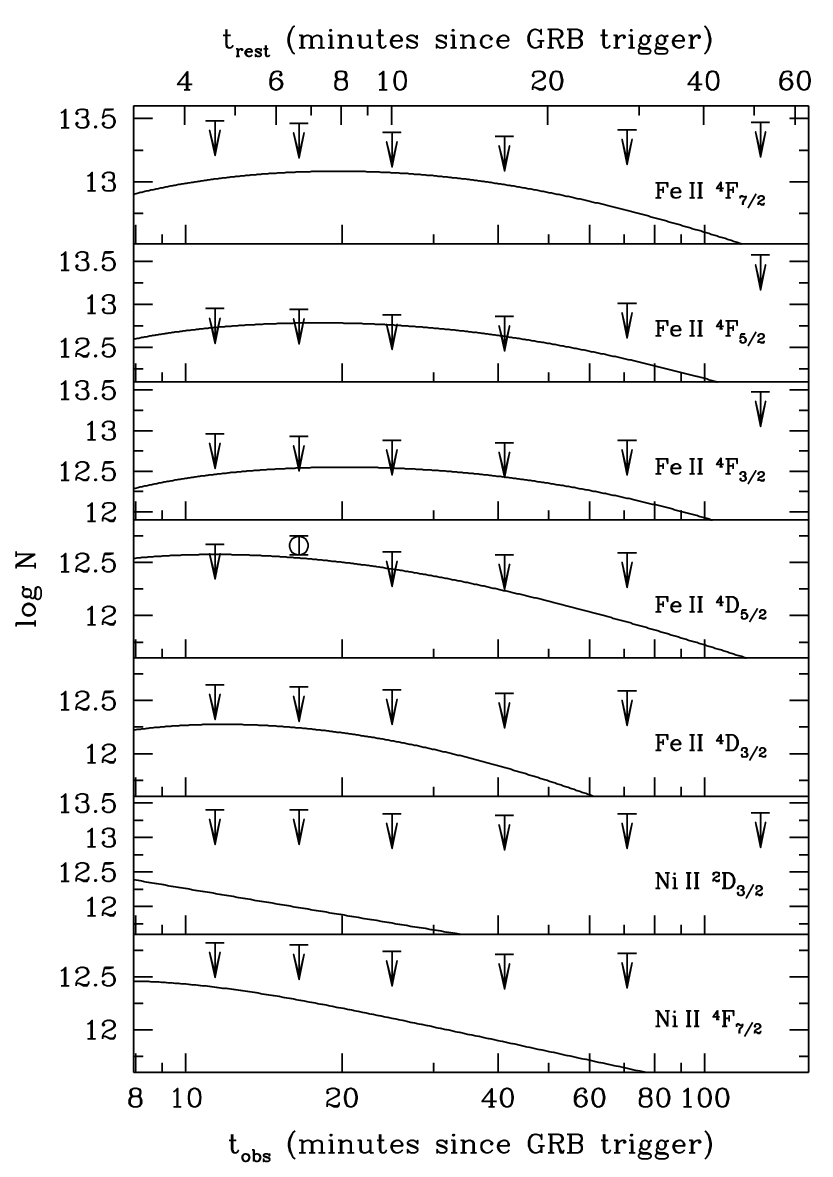}
  \caption{Comparison of the UV-model predicted column densities for
    several additional Fe~II and Ni~II excited levels, i.e. others
    than the ones shown in Fig.~\ref{fig:uv}, with the 5$\sigma$ upper
    limits that we obtain from the spectra. The Fe~II $^4$D$_{5/2}$
    level is actually detected at just above 5$\sigma$ in the second
    epoch. \label{fig:uvlimits}}
\end{figure}

As we calculate the level population of the lower 20 levels for Fe~II,
and lower 17 levels for Ni~II, we can compare if the model predictions
for all levels are consistent with our data. Searching for the
detections of lines originating from these levels has resulted in one
new detection (epoch 2 for level Fe~II $^4$D$_{5/2}$), but the rest we
can only place an upper limit (we adopt 5$\sigma$) to the column
density, as shown in Fig.~\ref{fig:uvlimits}.

\citet{2006astro.ph..6462D} have reported a significant
($\sim$3$\sigma$) decline by at least a factor of five in the
equivalent width of the Fe~II $^6$D$_{7/2}$ \la 2396 transition, in
spectra taken at 4.7 and 20.8~hours after GRB~020813. We note that
this line is saturated in our spectra, and moreover blended with Fe~II
$^6$D$_{5/2}$ \la 2396, and therefore we do not use it in our
analysis. To verify if this decline is roughly consistent with our
calculations for \grb, we determine the best-fit UV pumping model
Fe~II $^6$D$_{7/2}$ column densities for \grb\ at 2.1 and 9.2~hours in
the rest frame, using the redshift of GRB~020813, $z=1.255$
\citep{2003ApJ...584L..47B}. We obtain log~N(t$_{\rm
  rest}$=2.1~hr)=13.16 and log~N(t$_{\rm rest}$=9.2~hr)=12.45,
corresponding to a decrease of a factor of 5.1, fully consistent with
the result of \citet{2006astro.ph..6462D}.

\section{Discussion}
\label{sec:discussion}

Comparison of our UVES data with modeling of collisional and radiative
excitation clearly shows that UV pumping by GRB afterglow photons is
the mechanism responsible for the population of the Fe~II and Ni~II
excited levels. However, in the UV pumping model, one would expect the
ions in the ground state and the excited levels to be exactly at the
same location and velocity. As we discussed in
Sects.~\ref{sec:hostabsorbers} and \ref{sec:variability}, this is not
exactly the case. One simple way of resolving this apparent
discrepancy is to invoke more components than the three that we
resolve. A fourth and fifth unresolved component could be hidden in
the resonance-line profile at the same velocities as the two main
components of the excited levels. Then the observed velocity offset
between the excited levels and the resonance lines can be explained if
the observed resonance-line components are mainly due to gas that is
further away from the GRB, so excitation of the ground state by UV
pumping is negligible. This requirement of additional ground-state
components is consistent with the rather poor fit of our 3-component
Voigt-profile fit to the blue wing of a few high S/N resonance lines
(see Fig.~\ref{fig:sumprofiles}). The existence of additional
components is fully consistent with the UV pumping fit
results. According to this fit, the Fe~II and Ni~II ground state
column densities are 0.3-0.5 dex lower than what we measure in the
spectra, providing evidence for gas along the line of sight in the
host galaxy that is not affected by the GRB, and thus this gas needs
to be further away than the cloud that we modeled. Assuming that the
level ratio between the first fine-structure level and the Fe~II
ground state of this extra material is lower than 1/10, we estimate a
lower limit to its distance of $d=9.5$~kpc.

Our successful fit of the UV pumping model to the observed evolution
of the Fe~II and Ni~II excited levels has several interesting
implications.

The majority of the neutral gas closest to the GRB is at 1.7~kpc (but
see erratum in Appendix A). If there would have been neutral material
much closer in, we would not be able to reproduce the observed
evolution of the excited levels with our model. Naturally this value
is not accurate, mainly because of the uncertainty in the possible
extinction in between the GRB and the cloud. However, this extinction
is probably not very high. From the dust depletion pattern in the host
\citep[see Sect.~\ref{sec:hostabsorbers}
  and][]{2006astro.ph.11092P,2006NJPh....8..195S}, we estimate the
extinction to be low: A$_{\rm V}\sim$0.1. This value is in agreement
with the host-galaxy extinction estimate of A$_{\rm V}\sim$0.2 of
\citet{sara060418}, which is obtained from modeling the spectral
slope. Moreover, the spectral slope from the UV pumping fit,
$\beta=$\gpm{-0.5}{0.8}{1.0} is not very different from the observed
spectral slope after correcting for the extinction in the foreground
absorber at $z=1.1$ ($\beta=-0.8$), while a large amount of dust
extinction would severely affect the value of the spectral slope,
provided that the extinction is not grey.

We note that another lower limit to the absorber distance is set by
the presence of Mg~I \citep[see][]{2006astro.ph..1057P}, assuming that
it is at the same location as a large part of the Fe~II and Ni~II
excited material. For GRB~051111, \citet{2006astro.ph..1057P}
calculate that if Mg~I would be at a distance smaller than 80~pc from
the GRB, then Mg~I would have been fully ionized. Using Eq.~2 of
\citet{2006astro.ph..1057P}, we estimate the lower limit to the
distance where Mg~I can survive to be $d=45$~pc for \grb. More
recently, \citet{2006astro.ph.11079C} have estimated a lower limit to
the distance from GRB~021004 to the absorbers along its sightline that
are blue-shifted by 2500-3000~\kms. From the ratio of C~II\1star/C~II
they find $d>1.7$~kpc; limits for the other absorbers blue-shifted by
less than 2000~\kms\ are not given. A similar distance limit is set
for the Si~II gas associated with GRB~050730
\citep{2006astro.ph.11079C}. All these distance limit estimates are
fully consistent with our distance determination.

The consequence is that any pre-GRB neutral cloud that was present at
distances less than about 1.7~kpc (but see erratum in Appendix A), was
severely affected by \grb.  Atomic species typical of the neutral ISM
such as Fe~II, Cr~II, Zn~II, etc., are likely ionized to a higher
ionization stage. We note, however, that this does not imply that the
GRB has ionized all neutral material up to this distance; it may be
that the GRB ionized only its immediate surroundings, e.g. up to tens
of parsecs, and that between the ionized region and 1.7~kpc (but see
Appendix A), no significant amount of neutral material was
present. But it is clear that the immediate environment of GRBs cannot
be probed with these neutral ISM species, but possibly higher
ionization lines may be detected. It is therefore of great interest to
look out for higher ionization species not normally seen in optical
spectra, that may originate from the immediate surroundings of the
GRB. Possibly an ionization stratification could be observed, from
higher ionization lines close to the GRB, to lower ionization species
further out. X-ray spectroscopy instead of the optical will probably
be the best tool to probe the immediate vicinity of the GRB.

Because the photon energy threshold to ionize Fe~II to Fe~III is
higher than the ionization potential of H~I, this distance limit also
applies to neutral hydrogen, i.e. any significant H~I cloud that was
present before the GRB exploded at distances smaller than
approximately 1.7~kpc (but see Appendix A), will have been ionized. If
we assume that \grb\ is not special with respect to other GRBs in this
respect, most of the high-column density H~I clouds observed in GRB
afterglow spectra \citep{vrees030323,2006astro.ph..9450J} may also be
at typical kiloparsec distances. This assumption that \grb\ is not
special, is supported by the lower limit to the distances of absorbers
along the GRBs~021004 and 050730 sightlines determined by
\citet{2006astro.ph.11079C}, which are also of kiloparsec scale.
These H~I clouds could either be part of a giant star-forming region
in which the GRB was born, or simply clouds in the foreground in the
host galaxy. We note that if the H~I clouds are indeed at kiloparsecs
from the GRB, any metallicity estimate performed using optical
spectroscopy is most likely not representative of the metallicity of
the region where the GRB progenitor was born. The kiloparsec distance
for these absorbers, combined with the significant differences between
GRB-DLAs and QSO-DLAs in H~I column density
\citep{vrees030323,2006astro.ph..9450J}, metallicity
\citep{2006A&A...451L..47F} and dust depletion \citep[][ see also
  Sect.~\ref{sec:hostabsorbers}]{2006NJPh....8..195S}, suggests that
QSO sightlines are not probing the central kiloparsecs of (GRB)
star-forming galaxies. This is consistent with our observation in
Sect.~\ref{sec:hostabsorbers}, that the high-ionization profiles
follow those of the low-ionization species very well in \grb, while
this is uncommon in QSO-DLAs.

The gradual ionization of the neutral hydrogen close to the GRB could
be observed by monitoring the evolution of the \lya\ or metal line
\citep[][ we note that these authors suggest Mg~II as metal line
  probe, but this line is normally highly saturated, especially in the
  high density GRB host galaxy environments, and therefore not
  suitable for this purpose]{1998ApJ...501..467P}.  The redshift of
\grb\ is too low for \lya\ to be covered by our UVES spectra. From the
metal lines, we do not see any hint for such a gradual ionization;
very likely observations have to be performed even quicker than the
response time of our observations, to be able see this effect.

Our UV pumping fit shows that it is possible to obtain the distance of
the excited absorbing gas to the GRB. We have modeled the observations
with only one cloud, but this can be extended to a multiple cloud
model, where for each cloud one can obtain the distance with respect
to the GRB, its velocity, its Fe~II and Ni~II column density and the
cloud Doppler parameter, $b$. And for components not affected by the
UV photons, a lower limit to the distance can be set. This way it
could be possible to study the host-galaxy cloud structure, abundances
and kinematics in more detail than before.

Finally, the UV pumping fit not only constrains the properties of
clouds in the GRB host galaxy, but also two properties of the GRB
emission. One is the spectral slope of the GRB afterglow, even though
this value is not very tightly constrained in our fit:
$\beta=$\gpm{-0.5}{0.8}{1.0}. The second property is the total UV flux
that arrived at the cloud from the time of the burst trigger,
i.e. optical flash (if present) and afterglow combined. This flux can
be derived by determining the integral from fit parameter $t_0$ to any
time desired of the assumed light curve in the model. So even if no
UV/optical observations were performed by robotic telescopes or Swift
itself, the magnitude of a UV flash can be constrained from a UV
pumping fit.  For the case of \grb, we determine the limit on the
total observer's frame V-band flux that arrived from the GRB (UV flash
and afterglow) from the time of the GRB trigger to the start of our
first spectrum ($t$=11.4~minutes) to be
$(1.9\pm0.3)\times10^{-23}$~erg cm$^{-2}$ Hz$^{-1}$.  For comparison,
this flux is the same as that contained by a V=10 flash with a
duration of 5~seconds.

\section{Conclusions}
\label{sec:conclusions}

Using the VLT Rapid-Response Mode in combination with UVES, we have
obtained a unique time-series of high-resolution spectra of \grb\ with
a signal-to-noise ratio of 10-20.  These spectra show clear evidence
for variability of transitions arising from the fine-structure levels
of Fe~II, and from metastable levels of both Fe~II and Ni~II.  We
model the time evolution of the Fe~II and Ni~II excited levels with
three possible excitation mechanisms: collisions, excitation by IR
photons only, and UV pumping. We find that the collisional and IR
photon scenarios can be rejected. Instead, the UV pumping model, in
which a cloud with total column density $N$ and broadening parameter
$b$ at a distance $d$ from the GRB is irradiated by the afterglow
photons, provides an excellent description of the data. The best-fit
values are log~$N$(Fe~II)=\gpm{14.75}{0.06}{0.04},
log~$N$(Ni~II)=$13.84\pm0.02$, $d=1.7\pm0.2$~kpc (but see Appendix A),
and $b=25\pm3$~\kms. The main consequence of this successful fit, is
the absence of neutral gas, in the form of low-ionization metals or
H~I, at distances shorter than 1.7~kpc (but see erratum in Appendix
A).  Any pre-explosion neutral cloud closer to the GRB must have been
ionized by the GRB. Therefore, the majority of very large H~I column
densities typically observed along GRB sightlines may not be in the
immediate surroundings of the GRB; they could either be part of a
large star-forming region, or foreground material in the GRB host
galaxy. In either case, the metallicity derived from absorption-line
spectroscopy may not be representative of the metallicity of the
region where the GRB progenitor was born.

\begin{acknowledgements}
We are very grateful for the excellent support of the Paranal staff,
and in particular that of Stefano Bagnulo, Nancy Ageorges and Stan
Stefl. We have made extensive use of the atomic spectra database of
the National Institute of Standards and Technology (NIST), see
http://physics.nist.gov/PhysRefData/ASD/index.html.  This research was
supported by NWO grant 639.043.302 to RW.  This work has benefitted
from collaboration within the EU FP5 Research Training Network
``Gamma-Ray Bursts: an Enigma and a Tool''.

\end{acknowledgements}

\begin{appendix}

  \label{sec:appendix}
  
  \section{Erratum: Rapid-Response Mode VLT/UVES spectroscopy of \grb}

  We have recently realized that, in Eq.~3 in
  \citet{2007A&A...468...83V}, the flux $F_{\nu}(\tau_0)$ should be
  divided by $4\,\pi$.  The relation should therefore read as follows:
  
  \setcounter{equation}{2}
  \begin{equation}
    \frac{{\rm d}N_{\rm u}}{{\rm d}t} = N_{\rm l} B_{\rm lu}
    \frac{F_{\nu}(\tau_0)}{4\,\pi} - N_{\rm u} \left[A_{\rm ul}+B_{\rm ul}
      \frac{F_{\nu}(\tau_0)}{4\,\pi}\right]
    \label{eq:dndt}
  \end{equation}
  
  Using this relation, our excitation program is now fully consistent
  with the PopRatio code \citep[][see also
    Sect.~5.2]{2002MNRAS.329..135S} when neglecting collisional
  excitation, and in the optically thin regime as PopRatio assumes all
  transitions are optically thin. The consequence is that excitation is
  a factor of $4\,\pi$ less effective than we had previously assumed,
  resulting in a decreased distance estimate by a factor of
  $\sqrt{4\,\pi}\sim3.5$. Therefore, the distance of \grb\ to the
  neutral absorbing material -- previously $d=1.7\pm0.2$~kpc -- needs to
  be revised to $d=0.48\pm0.06$~kpc, under the same model assumptions.
  The main conclusion of the paper, that the neutral absorbing gas is
  not in the immediate environment of \grb, remains the same.
  
  Since we applied the same excitation analysis in Sect.~4.1.3 of
  \citet{2008A&A...491..189F} and in Sect.~2.3 of
  \citet{2009A&A...506..661L}, the distance estimates therein should
  also be scaled down by a factor of $\sqrt{4\,\pi}$. The main
  conclusions of these two papers are not affected by this change
  either.
  
\end{appendix}

\bibliographystyle{aa.bst}
\bibliography{references}

\end{document}